\documentclass{article}

\usepackage{arxiv}
\usepackage{float}
\usepackage{caption}
\usepackage[utf8]{inputenc} 
\usepackage[T1]{fontenc}    
\usepackage{hyperref}       
\usepackage{url}            
\usepackage{booktabs}       
\usepackage{amsfonts}       
\usepackage{nicefrac}       
\usepackage[disable]{todonotes}      
\usepackage{microtype}      
\usepackage{enumitem}
\usepackage{amsmath}
\usepackage{lipsum}		
\usepackage{graphicx}
\usepackage{natbib}
\usepackage{doi}
\usepackage{glossaries}
\usepackage{fontspec}
\usepackage{cleveref}
\usepackage{tikz}
\usetikzlibrary{trees}
\usetikzlibrary{arrows.meta,positioning,fit,calc,backgrounds,shapes.geometric}
\usepackage{unicode-math}

\newfontfamily\symbolfont[
  Path = Fonts/,
  Extension = .ttf
]{NotoSansSymbols2-Regular}

\newcommand{\AC}{{\symbolfont\char"10196\ }}
\newcommand{\ACEQ}{{\text{\symbolfont\char"10196}}}

\title{Liberata - Graph Scientometrics for a Share Based System of Academic Publishing}

\hypersetup{
pdftitle={Liberata - Graph Scientometrics for a Share Based System of Academic Publishing},
pdfsubject={},
pdfauthor={Han ~Zhang, Anshuman ~Sabath},
pdfkeywords={Liberata, Graph Theory, Scientometrics, Econometrics},
}

\begin{document}
    \author{
        \href{https://orcid.org/0000-0002-7522-8101}{\includegraphics[scale=0.06]{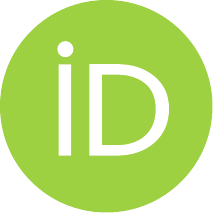}\hspace{1mm}Han Zhang$^{0.55}_{1}$}
        \quad
        \href{https://orcid.org/0009-0002-0326-1786}{\includegraphics[scale=0.06]{Figures/orcid.pdf}\hspace{1mm}Anshuman Sabath$^{0.30}_{2}$}
        \quad
        \href{https://orcid.org/0000-0002-9381-4630}{\includegraphics[scale=0.06]{Figures/orcid.pdf}\hspace{1mm}Timothy W. Dunn$^{0.05}_{2}$}
        \quad
        \href{https://orcid.org/0000-0003-2551-1563}{\includegraphics[scale=0.06]{Figures/orcid.pdf}\hspace{1mm}L. Catherine Brinson$^{0.10}_{1}$}
        \\[4pt]
        \texttt{han.zhang@duke.edu, anshuman.sabath@duke.edu, timothy.dunn@duke.edu, cate.brinson@duke.edu}
        \\[4pt]
        $_1$ Department of Mechanical Engineering \& Materials Science, Duke University, Durham, NC \\
        $_2$ Department of Biomedical Engineering, Duke University, Durham, NC \\
    }

\maketitle
\begin{abstract}

Contemporary scientometric indicators remain anchored in paradigms and axioms from when academic research was done in small scholarly communities. With the proliferation of professional scientific research around the globe, academic research is now mostly conducted in large communities, with high rates of information incompleteness about work impact and individual contributions. This has profound implications for how research output is measured and quality controlled, especially as the rate of academic publishing continues to climb. Exploits of complex systems are usually found at discrete transition points, where rules turn on or off, and academia is not immune to this natural pattern. Exploitative career boosting strategies are a well recognized and growing problem within academia, largely enabled by maligned incentive structures and traditional metrics forcing discretization of credit to authors (positions) and prior works (citations) when those are fundamentally continuous valued quantities.

This article introduces Liberata's scientometrics, a share-based framework for academic publishing and quality control. In this system, authorship positions are replaced with contribution shares, which sum to unity and give both ordinality and distances of contributions. These shares can be traded on Liberata's academic marketplaces for quality control services such as peer review and replication, rewarding quality controllers based on the long term success of the work. Citations are weighted to protect against frivolous citations and credit printing, and modular correction factors are introduced to allow multiple distinct measures of impact. Liberata's metrics are formalized through two fundamental graphs: 1.) Shares and 2.) References. From these two, the Liberata system constructs the notion of \emph{academic capital} and catalogs a naturally arising set of scientometrics that captures impact, risk, collaboration, collusion, value of quality control, diversification, etc. These metrics richly but succinctly represent academic contributions, not just for individuals, but are easily extensible to institutions, geographic regions, time periods, and research fields.

\end{abstract}

\keywords{Graph Theory \and Scientometrics \and Econometrics \and Academic Publishing \and Metascience}

\newpage
Author's note: This article is long because it is written to be more a catalog than a single narrative. Readers are encouraged to be selective in which sections they read after \cref{sec:introduction} and \cref{sec:liberata_system}.
\setcounter{tocdepth}{1}
\tableofcontents

\newpage
\section{Introduction}
\label{sec:introduction}
\todo{Fill in Introduction and flesh it out. Its in Notion -> Algo space -> Doc -> Literature Storage - HZ}

The measurement of scholarly impact is central to the evaluation and governance of science, influencing decisions related to hiring, funding allocation, and institutional ranking. Bibliometric indicators such as publication counts, citation counts, and composite indices including the h-index (Hirsch, 2005) remain widely used to quantify research performance. A substantial body of work has examined the properties and limitations of these indicators, including numerous variants designed to improve their sensitivity to citation distributions (Bornmann, Mutz, \& Daniel, 2008). Despite their prevalence, these indicators rely on simplified assumptions about the allocation of credit that are increasingly misaligned with the collaborative nature of modern scientific production.

A key limitation of existing systems lies in their treatment of authorship and citations. Most bibliometric indicators implicitly assume that publications and citations are indivisible units of credit that accrue equally to all listed coauthors. This assumption disregards the heterogeneity of contributions within collaborative research and obscures the internal structure of knowledge production. As research teams grow in size and complexity, these limitations become more pronounced.

Existing approaches have attempted to address this issue through alternative counting methods. Full counting assigns complete credit to each coauthor, leading to inflationary effects in multi-authored publications. Fractional counting mitigates this by dividing credit equally among authors, ensuring conservation of total credit, but retains the assumption of equal contribution (Waltman \& van Eck, 2015). Author-order–based weighting schemes provide a partial correction by assigning credit based on position, yet these conventions vary widely across disciplines and are often inconsistent or ambiguous (Waltman, 2012). As a result, these methods remain imperfect proxies for actual contribution.

More recent work has explored network-based approaches to modeling scholarly impact. Citation network methods, including PageRank-style algorithms, use the structure of citation graphs to estimate influence within the scientific literature (Chen, Xie, Maslov, \& Redner, 2007). Similarly, credit allocation models based on coauthorship and citation structures have been proposed to infer the distribution of credit among collaborators (Shen \& Barabási, 2014). While these approaches incorporate relational information, they typically infer contribution indirectly rather than representing it explicitly.

In parallel, initiatives such as the CRediT taxonomy have improved transparency in authorship by providing standardized descriptions of contributor roles (Allen et al., 2014). However, these frameworks are primarily descriptive and are not directly integrated into quantitative impact measures or citation-based indicators.

Taken together, these limitations point to a fundamental gap in current scientometric systems: the absence of a unified framework that simultaneously represents heterogeneous contributions and propagates impact through the scholarly network. Addressing this gap requires moving beyond discrete authorship conventions toward representations that encode contribution structure at a finer granularity.

In this paper, we introduce \emph{Liberata}, a share-based framework for modeling scholarly production and citation flows. In this system, contributors to a manuscript are assigned continuous contribution shares that sum to unity, replacing discrete authorship ordering. These shares define a weighted representation of the scholarly ecosystem in which citations propagate as impact signals through a network of manuscripts and contributors. The interaction between contribution shares and citation flows yields a contribution-weighted measure of impact, termed \emph{academic capital}.

Building on this framework, we develop a set of scientometric indicators that capture not only aggregate impact but also its distribution across contributors and roles. These include measures of contribution concentration, the valuation of peer review and replication activities, and the dynamics of impact accumulation within the system. By integrating contribution shares with network-based citation analysis, the proposed approach provides a coherent and extensible foundation for contribution-aware scientometrics.

The remainder of the paper formalizes the Liberata framework, introduces the underlying graph representations, and develops the associated metrics for analyzing scholarly impact within this share-based system.

\newpage
\section{Liberata System of Open Access Publishing}
\label{sec:liberata_system}
In Latin, "liber" means "book" or "free", and "rata" means "rate" or "ratio". The two together can be interpreted to either mean free [market] rate, or publication shares. The fundamental idea of the Liberata open access publishing system is to use shares to denote paper contributions and to allow the trading of these shares for quality control services (peer review \& replication). With the share-based system demonstrated in \cref{fig:liberata_system}, the aspiration is to liberate academia from the perverse game theory incentives and subsequent politics and rent seeking problems of the traditional academic publishing system and allow for a self-sufficient system of accurate work accreditation, incentivized quality control, open access publishing system.

\subsection{Shares Based Contribution Attribution} 
\label{ssec:shares_system}

\subsubsection{Traditional Contribution Attribution}
In traditional research manuscripts, work is recognized through the author list, of which there are some common conventions to interpret the positions of the authors.
\begin{itemize}
    
    \item \textbf{Descending order:} In most fields of academic research, the authors are listed in descending order of perceived contribution to the work with the exception that the last author position is reserved for the principal investigator or supervisor of the project \cite{Shamoo2009,Sauermann2017}.
    
    \item \textbf{Co-first author:} this convention is used in fields where many different types of key contributions are often made to a work, such as microbiology. This practice attempts to allow multiple people to get recognition on academic metrics that count only or differently first authors from other authors \cite{Riesenberg1990,Tscharntke2007}.
    
    \item \textbf{Alphabetical order:} this convention is present in economics and mathematics, whereby there is no association between author position and contribution \cite{Einav2006,Waltman2012}. This practice is used in fields where it is very subjective or difficult to rank order authors' contributions. In such subjective cases, discussions about rank ordering, particularly if collaborations are in small groups and likely to be repeated for many projects, impose a substantial social cost.
    
    \item \textbf{Reverse ordering of supervisors:} in some fields, the supervisors of a project (professors) are listed in reverse order of contribution where the main supervisor is at the end \cite{Tscharntke2007,Bennett2003}, and other supervisors in ascending order of contribution prior to the main supervisor. This is a practice seen in fields where it is common to have multiple professors advising on a project, and there is a need for more than the last author position to indicate that these are supervisors. Readers, upon considered inspection, could work out who is a student/worker with relatively minor contribution versus a professor/supervisor with relatively high contribution.
    
    
    
    
    \item \textbf{CRediT Roles:} The CRediT taxonomy \cite{Allen2014} provides more context about the types of contributions different authors have made to the paper. However, this still does not reveal how much effort was put in for each of the contributions, and how the rest of the authors value such contributions. This disconnect opens up avenues for authorship politics still. Further the roles do not provide any mechanism to include contributions from non-authors, e.g., peer reviewers and other entities who put in valuable work in making a paper higher quality.
    \todo{Done: Fill up - AS}
\end{itemize}

The above conventions for recognizing contributions is a problem within academia. 
The fundamental issue with the traditional system is that authorship positions are a discrete system of recognition trying to track contributions that are inherently continuous, which inevitably leads to problems of unfair or inaccurate credit assignment \cite{Tscharntke2007,Sauermann2017}. In general, there is no agreement between scholars given an author list how precisely
credit should be distributed between the authors \cite{Shen2014,Hagen2008}.

\subsubsection{Contribution Shares in Liberata}
Liberata swaps out the discrete authorship position credit system with a continuous system of contribution shares which denote percentage contribution to a manuscript. Every person on the author list is assigned a percentage contribution to a manuscript upon submission to the Liberata platform. Subsequently, other contributors such as peer reviewers or replicators may come to hold shares on the manuscript for their services to the manuscript. Let $C_m = \{c_{1}, c_{2}, \dots, c_{n}\}$ be the set of $n$ contributors on a manuscript $m$, and let $S_m = \{s_{m,1}, s_{m,2}, \dots, s_{m,n}\}$ be the set of shares held by those contributors on that manuscript. By construction, contribution shares have the following properties: 

\begin{itemize}
    \item Shares for any contributor $c$ (from authors $A_m$, peer reviewers $P_m$, and replicators $R_m$) on any manuscript $m$ must be denoted as a real number between 0 and 1.
    \begin{equation}
        s_{m,c} \in [0,1] \ \forall \ c \in C_m = A_m \cup P_m \cup R_m     
            \label{sharesValidValues}
    \end{equation}
    \item The sum of contribution shares for all people involved on a paper for any manuscript is unity. 
    \begin{equation}
        \sum_{k=1}^{n} s_{m,k} = 1
            \label{sharesSumUnity}
    \end{equation}
    \item The value (academic capital) of a contributor's shares on a manuscript $\ACEQ_{m,c}$ is equal to the product of share of that manuscript and the (weighted) citations $w$ of that manuscript.
    \begin{equation}
        \ACEQ_{m,c} = s_{m,c} w_m    
            \label{sharesAC}
    \end{equation}
    \item All shares are fungible for the same manuscript.
    \begin{equation}
        \ACEQ_{m,c1} = \ACEQ_{m,c2} \iff s_{m,c1} = s_{m,c2}    
            \label{sharesFungible}
    \end{equation}
\end{itemize}

Prior work on co-authorship credit allocation has proposed a range of weighting schemes, including equal fractional counting \cite{Egghe2008,PerianesRodriguez2016}, author-order-based heuristics \cite{zhang2009proposal,Howard2007}, and inferred contribution models based on citation structure \cite{Shen2014}. 
More recent work has emphasized the multidimensional and continuous nature of contributions \cite{Sauermann2017,Allen2014}, though most systems stop short of assigning quantitative shares.

The contribution share system in Liberata is closely related to fractional and contribution-based approaches in that it assigns continuous credit weights that sum to unity across contributors. However, it differs in a key aspect - \textit{contribution shares are treated as explicit primitives rather than inferred quantities}. This allows the framework to generalize beyond authors to include other contributors such as peer reviewers and replicators. As such, the proposed system can be viewed as a natural generalization of existing co-authorship weighting schemes within a unified capital allocation framework.

\begin{figure}[H]
\centering
\begin{tikzpicture}[
    font=\small,
    >=Latex,
    node distance=0.9cm and 1.2cm,
    rolebox/.style={
        draw,
        rounded corners=2pt,
        minimum width=2.6cm,
        minimum height=0.9cm,
        align=center,
        fill=gray!8
    },
    paperbox/.style={
        draw,
        rounded corners=2pt,
        minimum width=2.8cm,
        minimum height=1.0cm,
        align=center,
        fill=gray!15
    },
    sourcebox/.style={
        draw,
        rounded corners=2pt,
        minimum width=2.4cm,
        minimum height=0.85cm,
        align=center,
        fill=gray!5
    },
    metricbox/.style={
        draw,
        rounded corners=2pt,
        minimum width=3.5cm,
        minimum height=1.0cm,
        align=center,
        fill=gray!10
    },
    groupbox/.style={
        draw,
        rounded corners=4pt,
        inner sep=8pt
    },
    shareedge/.style={-{Latex[length=2.2mm]}, semithick},
    citedge/.style={-{Latex[length=2.2mm]}, semithick, solid},
    flowedg/.style={-{Latex[length=2.4mm]}, thick},
    note/.style={align=center},
]

\node[rolebox] (author) {Author(s)\\$a \in A$};
\node[rolebox, below=0.3cm of author] (reviewer) {Peer Reviewer(s)\\$p \in P$};
\node[rolebox, below=0.3cm of reviewer] (replicator) {Replicator(s)\\$r \in R$};

\node[paperbox, right=3.5cm of reviewer, minimum height=2.5cm] (paper) {Manuscript\\$m \in M$};

\node[sourcebox, right=3.5cm of paper] (cite2) {Citing paper \\ $\ell \in M$};
\node[sourcebox, above=0.3cm of cite2] (cite1) {Citing paper \\$k \in M$};
\node[sourcebox, below=0.3cm of cite2] (cite3) {Citing paper \\$j \in M$};

\draw[shareedge, dashed, -] 
($(paper.north west)!0.25!(paper.south west)$) -- 
node[above, sloped, pos=0.5] {$\ACEQ_{m,a} = s_{m,a} * \ACEQ_m$} 
(author.east);

\draw[shareedge, dashed, -] 
(paper.west) -- 
node[above, sloped, pos=0.5] {$\ACEQ_{m,p} = s_{m,p} * \ACEQ_m$} 
(reviewer.east);

\draw[shareedge, dashed, -] 
($(paper.north west)!0.75!(paper.south west)$) -- 
node[above, sloped, pos=0.5] {$\ACEQ_{m,r} = s_{m,r} * \ACEQ_m$} 
(replicator.east);

\draw[citedge] (cite1.west) -- node[above, sloped] {$w_k = 1/|\mathrm{ref}|_k$} ($(paper.north east)!0.25!(paper.south east)$);
\draw[citedge] (cite2.west) -- node[above] {$w_l = 1/|\mathrm{ref}|_\ell$} ($(paper.north east)!0.5!(paper.south east)$);
\draw[citedge] (cite3.west) -- node[above, sloped] {$w_j = 1/|\mathrm{ref}|_j$} ($(paper.north east)!0.75!(paper.south east)$);

\node[fit=(author)(reviewer)(replicator), inner sep=0pt] (groupL) {};
\node[fit=(paper)(cite1)(cite2)(cite3), inner sep=0pt] (groupC) {};

\node[note, below=0.7cm of paper, text width=10.8cm] (captionnote)
{Weighted citations from citing papers determine manuscript impact $\ACEQ_m= w_m =\sum 1/|\mathrm{references}|$, which is then allocated across author, reviewer, and replicator roles in proportion to their shares in $m$.};

\end{tikzpicture}
\caption{Schematic of the Liberata framework. Contributor-role nodes hold manuscript shares \(s_{m,c} \quad c\in C \) in the shares graph, while weighted citations define manuscript impact \(w_m\) in the references graph. Their composition yields manuscript-level academic capital allocations \(\ACEQ_{m,c}=s_{m,c}\ACEQ_m\) across authors, peer reviewers, and replicators.}
\label{fig:liberata_system}
\end{figure}

\subsection{Weighted Citations \& Correction Factors}
\label{ssec:weighted_citations}
\subsubsection{Traditional Citations}
Traditional citations, which Liberata refers to as absolute or unweighted citations $u_m$, counts the number of times a paper $m$ appears in the references sections of later papers. This is a proxy measure for how "impactful" the paper is to future work. This is the primary way the impact of manuscripts are measured in academia at present. Suppose $M'$ represents the manuscripts published after $m$ and the ref$(m',m)$ function returns 1 if $m'$ has $m$ in its references section, and 0 otherwise. Unweighted citations are then defined as:
\begin{equation}
    u_m = \sum_{M'}\text{ref}(m',m)
    \label{citationsUnweighted}
\end{equation}

This method of calculating impactfulness of a work is highly problematic for the following reasons.

\begin{itemize}
    \item Citations do not indicate endorsement. Citations can indicate something is being refuted, criticized, or corrected. Yet regardless of whether a work is positively influential to another work, or completely negated by another work, that earlier work gets the same amount of credit.
    \item Citations are not normalized across disciplines. Different fields of academia differ substantially on typical size of references section, leading to high distortions in perceived impactfulness in different fields and especially for interdisciplinary scholarship.
    \item Citations are not normalized for the number of works in a reference section. This gives each paper published the ability to print unlimited credit into the universe and leads to the steady inflation of reference sections and frivolous citations.
\end{itemize}

While many of these caveats have been recognized in prior works, no concrete solution has emerged addressing all the shortcomings of impact quantification. 
Prior citation network based analytical works have explored different methods for weighting the edges to extract different metrics of interest and investigate network effects. For instance, bibliographic coupling \cite{Kessler1963} and co-citation \cite{Small1973} weightings construct similarity measures between documents based on shared references or joint citation patterns, and are widely used to uncover topical proximity and clustering structure in the literature. Derivatives of PageRank applied to citation graphs \cite{Page1999,Chen2007} instead emphasize global importance by recursively weighting citations according to the influence of the citing sources. Age-based weighting schemes \cite{Walker2007} incorporate temporal decay to privilege recent contributions or model citation dynamics, while field-normalized metrics \cite{Waltman2011} attempt to correct for systematic differences in citation practices across disciplines.

Despite their utility, these weighting schemes exhibit several limitations. 
Similarity-based methods such as bibliographic coupling and co-citation are inherently static and can overemphasize well-established areas, failing to capture emerging or interdisciplinary connections \cite{ Porter2009, Wagner2011}. 
PageRank-type approaches introduce rich-get-richer dynamics, which can amplify early advantages and entrench already prominent works or authors. 
Age-based weightings are sensitive to the choice of decay function and may undervalue foundational contributions \cite{Wang2013,Parolo2015}. 
Field-normalized methods, while addressing cross-disciplinary heterogeneity, rely on often loosely-defined or externally imposed field classifications, which can be coarse, overlapping, or manipulable \cite{Waltman2013,Glanzel2010}. 
Moreover, the weighting methods explored in the prior works are susceptible to being used by citation cartels \cite{Fister2016} to artificially bolster citation counts on their own articles. 
This problem has already been emerging in current systems of peer-reviewing \cite{Thurner2011,Szomszor2020}. 
More broadly, many of these approaches lack explicit guardrails against cumulative advantage, leading to reinforcement of existing hierarchies and limiting their ability to fairly represent contribution in evolving scientific landscapes.

\subsubsection{Liberata's Weighted Citations}
\label{sssec:liberata_weighted_citations}
The Liberata system introduces weighting mechanisms with corrections which, together, tackle all the listed problems of impact quantification. By default, Liberata system uses weighted citations $w$ which normalizes the credit coming from each citation by the number of works cited in the referring paper, thus limiting the credit each paper can print from an uncapped citation number to 1 unit citation. 
\begin{equation}
    w_m = \sum_{k=1}^{n}\frac{1}{|\text{ref}|_k}
    \label{citationsWeighted}
\end{equation}

In addition to a normalization for reference count, there are two other optional correction factors that are applied by default in the Liberata system but can be toggled off or thought of as a variant of default metrics. 
\begin{itemize}
    \item Normalization for the publication rate of the academic field of the work. Suppose for a given academic field $d_4$ the average publication rate is $\rho(d_4)$ manuscripts per year per person. Then, we normalize all weighted citations $w$ coming from manuscripts $m \in M_{d_4}$ to 
    \begin{equation}
        w' = \frac{w}{\rho(d_4)}
            \label{citationsPublicationRate}
    \end{equation}
    This implies that the Liberata system considers all academic fields equally productive on average, so that the amount of unit citations and \AC generated in each field per researcher is the same per unit time.
    \item Normalization for the author similarity of cited works. Suppose $m'$ cites $m$. Let $\bar{v}_{m'}$ and $\bar{v}_{m}$ be vectors representing the share split amongst authors of the manuscript. The similarity is defined as:
    \begin{equation}
        \phi_{m',m} = \bar{v}_{m'} \cdot \bar{v}_{m}
            \label{authorSimilarity}
    \end{equation}
    The weighted citation going from $m'$ to $m$ is scaled by the author dissimilarity between the two works.
    \begin{equation}
        w'_{m',m} = (1-\phi_{m',m})w_{m',m}
            \label{citationsAuthorSimilarity}
    \end{equation}
    This correction factor is meant to nullify citation value of self accreditation.
\end{itemize}

\subsection{Academic Capital}

Academic capital (\AC) is a core concept of the Liberata system. \AC is a measure of impact and contributions to academia and is defined over some set of papers $M$ and shares held on $M$ by people $C$. 
\begin{equation}
    \ACEQ_{M,C} = \sum_{m\in M, c \in C} s_{m,c}\cdot w_m
    \label{ACDefinition}
\end{equation}

\AC is well defined over multiple situations. If we restrict $C$ to a single contributor "$c$", then $\ACEQ_c$ represents $c$'s career academic contributions across all academic fields. If we take $C$ to be a set of individuals from a lab or institution, the impact and productivity of that lab or institution can be quantified in a way H-index \cite{Hirsch2005} cannot generalize to (due to ill defined double/triple/... counting of citations). If we do not restrict $C$ and instead restrict $M$ to only be papers from certain fields or certain time periods, we can measure the volume of contributions within different fields of academic research or during certain time periods, or both. The versatility of defining subsets of $M$ and $C$ to examine allows for very simple yet powerful measures into the productivity and impact of different people, places, times, and research fields, from something as granular as the contributions of female authors from a particular institution during a period of time for a particular academic field, to something as broad as the total academic output of STEM fields globally per generation.

In addition to versatility, \AC is a superior metric to the h-index and citation count because it has a higher resolution while being a more accurate measure of individual contributions. This is possible because \AC has built in information not just of the ordinality of contributions, such as authorship position, but also of the distances in contribution, something traditional metrics do not have and cannot infer. Additionally, the normalizations and corrections of the citations allow for consistent interpretations across academic fields and time periods of the same numeric quantity. 

\subsection{Marketplaces}
\label{ssec:marketplaces}
With Liberata's shares-based accreditation system, it is possible to recognize, in addition to the contributions of authors, those of other crucial players in the scientific process. Specifically, on the two marketplaces of the Liberata open access publishing platform, peer review and replication services, respectively, can be purchased by authors with contribution shares. 

In the peer review marketplace, the authors can place a bid for peer reviewer services, specifying the number of peer reviews desired and percentage shares for each peer reviewer. To the authors, if the post-peer review manuscript $m'$, (on which the authors would have fewer shares but expect to accrue more lifetime citations), is likely to have higher academic capital than the pre-peer review manuscript $m$, then the authors are incentivized to go through peer review because the process has an expected net benefit on their conventional and new academic metrics (detailed in sections below). The equation below represents that condition for transaction.
\begin{equation}
    \begin{aligned}
    \mathbb{E}[\ACEQ_{m',A}(s_{m',A}, \mathbb{E}[w_{m'}])] > \mathbb{E}[\ACEQ_{m,A}(s_{m,A}, \mathbb{E}[w_m])] \\
    \text{where } s_{m',A} < s_{m,A} \text{ and } \mathbb{E}[w_{m'}] > \mathbb{E}[w_m]
    \end{aligned}
    \label{eq:condition_for_transaction}
\end{equation}

Alternatively, because \AC is proportional to $s$ and $w$, this condition can also be expressed as:
\[E[w_{m'}] > \frac{s_{m,\ A}}{s_{m',\ A}}E[w_m]\]
Conversely, if the above were not true, the authors would not be incentivized to go through the peer review process and could leave their manuscript on Liberata as an unreviewed manuscript which is still readable to all and equivalent to the preprint state of manuscripts today on open access platforms such as arXiv. 

For the peer reviewer $p$, it is worth taking $s_{m',p}$ to do this peer review if the time taken for the peer review $t_p$ is less than $s_{m',p}$ of what it would take to research and author a similar quality work (measured by $E[\ACEQ]$) by themselves $t_a$. 
\[t_p < t_a \cdot s_{m',p}\]

If both of these conditions are met, a transaction can be facilitated by a platform such as Liberata to allow for peer review to happen where both sides are incentivized and interest aligned to do good faith quality control. Unlike in other peer review systems, here the peer reviewers could expect to see greater returns if the work they reviewed actually ended up being better written or more accurate \cite{}. 

In the replication marketplace, the authors can place a bid for replication services, specifying the number of replicators desired, the percentage shares given to each successful replicator, and the time allotted for replication. A similar logic exists for authors on this marketplace as compared to the peer review marketplace, where the authors are comparing the expected academic capital of the fewer author shares on the post-replication paper $m''$ to the greater author shares on the pre-replication (post-peer review) paper $m'$. If the following is true, it is worth it for the authors collectively to pursue replication services.
\[E[w_{m''}] > \frac{s_{m',\ A}}{s_{m'',\ A}}E[w_{m'}]\]
Simutaneously, the replicator is comparing the time it would take to do the replication study on $m'$ compared with the opportunity cost of working on their own paper. The condition they need satisfied for the replication to be worth doing is.
\[t_r < t_a \cdot s_{m'',p}\]

Like with the peer review marketplace, if both of these conditions are met, a transaction can occur, allowing for a replication study to happen where the replicator is incentivized to do good faith quality control and is interest aligned with the author. Here, replicators could expect to see greater returns on their contribution shares if the work they replicated actually ended up with a revised, more accurate conclusion, or was replicated, verified, and thus more trustworthy.

\subsection{Academic Graphs} \label{ssec:academic_graphs}

Graphs are a foundational tool in scientometrics, used to represent and analyze the structure of scholarly communication \cite{Fortunato2018}. 
Traditionally, the academic ecosystem is typically modeled as a directed graph in which nodes correspond to entities such as manuscripts, authors, or journals, and edges encode relationships such as citation, co-authorship, or institutional affiliation \cite{Newman2010}.

The most common representation is the citation network, where each node represents a manuscript and a directed edge from $m'$ to $m$ indicates that $m'$ cites $m$ \cite{Kessler1963,Small1973}. 
This structure underlies a wide range of influence metrics, including citation counts, PageRank-style centrality measures \cite{Page1999,Chen2007}, and various field-normalized impact indicators \cite{Waltman2011,Waltman2015}. 
Similarly, co-authorship graphs model collaborations by connecting authors who have jointly produced a manuscript, often yielding insights into community structure and knowledge diffusion \cite{Newman2001,Barabasi2002}.



While these graph-based approaches have proven useful in many applications, they exhibit two structural limitations (below) that impede conventional scientometrics in accuracy and precision.

\begin{itemize}
    \item \textbf{Boolean edge semantics:} Relationships graphs such as citation and coauthorship networks are typically encoded as boolean edges (present or absent), without capturing the intensity, quality, or context of the interaction. In these cases, authorship contributions and citations implicitly are treated as equal, leading to many distortions and major exploits in downstream scientometrics.
    
    \item \textbf{Node homogeneity:} Many models treat nodes within a graph as belonging to a single type (e.g., manuscripts in citation networks, authors in coauthorship networks), or rely on loosely coupled multi-graph representations that do not fully integrate different roles (authors, reviewers, replicators) into a unified framework. This results in missing interactions between different types of objects in academic ecosystems, and blindspots in downstream scientometrics.
\end{itemize}

Liberata addresses these limitations by constructing two continuous valued graph representations: the \textit{Shares Graph}, and the \textit{References Graph}.

\paragraph{Shares graph (Section \ref{sec:shares_graph}).}
The shares graph encodes the relationship between contributors and manuscripts through contribution shares $s_{m,c}$. In contrast to coauthorship graphs, which represent collaboration as a boolean edge, the shares graph assigns a continuous weight to each contributor--manuscript relationship. This directly resolves the loss of contribution resolution in traditional models by making the magnitude of participation explicit. Moreover, all contributors—including authors and other roles are represented within the same graph, avoiding the fragmentation induced by node-type separation.

\paragraph{References graph (Section \ref{sec:references_graph}).}
The references graph represents directed relationships between manuscripts, where edges correspond to citations. Unlike traditional citation networks, these edges are not treated as uniform: their effect is scaled for number of references by default and optionally by other correction factors in section \ref{sec:exploits_&_modifications}. This standardizes the credit each manuscript prints into the universe, immediately removing some exploitative behaviors (e.g. frivolous citations) while also providing a more accurate accounting of manuscript impact.

With these two graphs, many other information rich representations can be constructed by products and powers of the two. A particularly useful one is the capital graph (Section \ref{ssec:capital_graph}), which combines the shares graph and citations graph to show the allocation of academic capital for all contributors. The details of each graph are described in the following sections. 
\todo{DONE: P1: redo capital graph intro after section 5 is reworked to just be about capital graph - HZ}

\newpage
\section{Shares Graph}
\label{sec:shares_graph}
\todo{More gradual introduction, perhaps in Section 2. Improve coherence - HZ}
\todo{P4: Read through for coherence, notation consistency, correctness  - AS}
The shares graph $G_S = ([C,M],S)$ used in the Liberata system stores all contributors $C$ and manuscripts $M$ as nodes. Edges $S$ of $G_S$ represent the contribution shares $c \in C$ holds in $m \in M$. The elements of $C$ are further subdivided into authors nodes $a \in A$, peer reviewer nodes $p \in P$, and replicator nodes
$r \in R$. Thus, there are $3c$ nodes person nodes denoting three possible roles academic contributors can hold on any manuscript. The shares held by each role for each person are separately recorded for all manuscripts. (Note: for any one manuscript, a person can only be either an author, peer reviewer, or replicator, and never more than one of these roles.)

$G_S$ has the following properties:
\begin{itemize}
    \item Each person is represented by 3 nodes (author, reviewer, replicator). $|A| = |P| = |R| = \frac{1}{3}|C|$
    \item All edges are from $c$ nodes to $m$ nodes. $S(M,M) = S(C,C) = \emptyset$
    \item All edge weights are non-negative. $\forall s \in S, \space s \geq 0$
    \item All the edges touching each $m$ node sum to 1. $\forall m \in M, \space deg(m) = \sum_{c \in C} s_{c,m} = 1$
\end{itemize}

The graph representation $S$ would comprise of the following list of nodes $M, A, P, 
R$, where the contributor nodes are duplicated such that the manuscript nodes $M$ form
a bipartite graph with each of the other groups, i.e., $(A,M)$, $(P,M)$, $(R,M) \subset S$ are 
three bipartite graphs, representing different roles of the contributors, which would form subgraphs of the graph $S$. 
Such a representation would yield an undirected graph and allow computation of several graph-based metrics 
which we would discuss in the subsequent sections.

\subsection{Matrix Representation}
\label{ssec:matrix_representation}
The adjacency matrix of $G_S$, $\mathbf{G_S}$, has a (bipartite) block structure below.
\begin{equation}
\mathbf{G_S} =
\left[
\begin{array}{c|c|c|c}
\mathbf{0} & \mathbf{S_A} & \mathbf{S_P} & \mathbf{S_R} \\
\hline
\mathbf{S_A}^T & \mathbf{0} & \mathbf{0} & \mathbf{0} \\
\hline
\mathbf{S_P}^T & \mathbf{0} & \mathbf{0} & \mathbf{0} \\
\hline
\mathbf{S_R}^T & \mathbf{0} & \mathbf{0} & \mathbf{0}
\end{array}
\right] \label{eq:shares_matrix_form}
\end{equation}
\todo{Done: Talk about the condensed shares graph - HZ}
$\mathbf{G_S}$ is a sparse matrix, with 10/16 blocks being $\mathbf{0}$, and the remaining 6/16 blocks carrying edge information about the shares authors $A\subset C$, peer reviewers $P \subset C$, and replicators $R \subset C$ carry in $M$. By construction, some useful properties of $\mathbf{G_S}$ are:
\begin{itemize}
    \item $\mathbf{G}_S$ is symmetric, i.e. $\mathbf{G}_S = \mathbf{G}_S^T$
    \item $\mathbf{G}_S = \mathbf{Q}_S \boldsymbol{\Lambda}_S \mathbf{Q}_S$ where the columns of $\mathbf{Q_S}$ are the orthonormal eigenvectors of $\mathbf{G_S}$ and $\boldsymbol{\Lambda_S}$ is a diagonal matrix with entries being the eigenvalues of $\mathbf{G_S}$
    \item All eigenvalues of $\mathbf{G_S}$ are guaranteed to be real numbers $\in [-1,1]$ with the largest eigenvalue guaranteed to be 1.
\end{itemize}
The first property is useful for analysis of academic capital in following sections, and the eigenvector and eigenvalue properties are useful for graph spectral analysis and clustering algorithms.

Note, this matrix would be very sparse, and it would be stored using standard best practices for storing sparse matrices. Throughout the rest of the paper, we still use the expanded form for better readability.

\subsection{Fetch Vectors}
\label{ssec:fetch_vectors}
When $\mathbf{G_S}$ is multiplied by a unit vector, the resulting vector can be interpreted as a distribution of shares. If the unit vector $v_m$ is along a dimension corresponding to a manuscript $m \in \mathbf{M}$ index, then $\mathbf{G_S}v_m=\mathcal{d}_m$ gives the distribution of shares for every author, peer reviewer, and replication involved in $m$. If instead the unit vector $v_c$ is along a dimension corresponding to a contributor $c \in A \cup P \cup R$, then $\mathbf{G_S}v_c=\mathcal{d}_c$ which corresponds to the shares a contributor holds as an author, peer reviewer, or replicator, across all papers respectively. 

Thus, unit vectors can serve as fetching mechanisms to return the share distribution across (1.) all contributors on a manuscript, or (2.) all manuscripts for a person for a given role. As will be seen in later sections, these distributions will have great use in calculating individual, institution, field, and time period metrics.

\subsection{Compositions of Fetch Vectors}
\label{ssec:compositions_fetch_vectors}
We can take a superposition of unit vectors to fetch more complex distributions of interest. For example, to find a person's total contributions across all roles and all papers, we construct $v'_{c} = v_{a}+v_{p}+v_{r}$ where $a$ is this contributor's author index, $p$ is their peer reviewer index, and $r$ is their replicator index, all of which are different dimensions in the matrix space by construction. The resulting $v'_{c}$ thus has three $1$ entries and all other entries $0$. 
\todo{P3:Equations should be on their own line - HZ}
$\mathbf{G_S}v'_c = \mathcal{d'_c}$ gives the contributor's distribution of shares across all papers, regardless of their type of contribution. Similarly, picking a $v$ corresponding to a subset of all authors, reviewers, replicators, or contributors allows for fetching of the shares held by entire labs, institutions, nations, etc. If we take a composition vector $v'_{M'} = \sum_{m \in M' \subseteq M}v_m$ across a subset of manuscripts, we can find from $\mathbf{G_S}v'_{M'} = \mathcal{d'}_{M'}$ the distribution of contributions across the academic community for a field of study, a particular time period, or for manuscripts arising from a certain geographical region or institution.

\subsection{Basic Distribution Metrics}
\label{ssec:basic_distribution_metrics}
Given a fetched distribution $\mathcal{d}=\mathbf{G_S}v$, there are medley possible statistical metrics that can examined to discern meaningful interpretations. For brevity, we limit the discussion in this section to just the common statistical moments (mean, variance, skew) as well as the median, mode, max, and min. Some examples of what could be measured are given below for inspiration, as an exhaustive list would be too lengthy to compile.

Suppose the distribution fetched $d$ was shares on a set of manuscripts for an individual. The mean of $d$ gives a sense of the typical scale of contributions for this contributor. The standard deviation tells us whether the individual consistently contributes at that level or plays a varied mix of minor and major roles on projects. Any abnormally common modes detected could reveal abnormal power dynamics or policies at play.

Suppose the fetched $d$ was for shares held by multiple contributors on a manuscript. The median of $d$ compared with the mean indicates whether most authors have minor or major roles. The standard deviation tells us the variation of contributions from each individual. If the collection of papers were for a topic, or academic field, one could see whether the field is equally distributed in expertise, and either it is new, growing, or shrinking. 

Suppose $d$ is for shares held by institutions on papers of an entire academic field for some period of time. One could find the concentration of research output for these this field during an era amongst all institutions, as well as discern what new versus mature institutions look like in research output. 

\subsection{Degree Matrix \& Laplacian Matrix}

Let $\mathbf{D_S}$ be the degree matrix of $G_S$ with elements defined as:

\[
d_{ij}  =
\begin{cases}
\sum_{k \in G_S} s_{i,k}, & \text{if } i = j \\
0 & \text{if } i \ne j
\end{cases} 
\]

\begin{equation}
\mathbf{D_S} =
\left[
\begin{array}{c|c|c|c}
\mathbf{I_{|M|}} & \mathbf{0} & \dots & \mathbf{0}\\
\hline
\mathbf{0} & \ddots & \ddots & \vdots \\
\hline
\vdots & \ddots & d_{k-1} & 0 \\
\hline
\mathbf{0} & \dots & 0  & d_{k}
\end{array}
\right]
\end{equation}

$\mathbf{D_S}$ only has nonzero elements along its main diagonal and those numbers represent total shares connected to the node. Thus for indices corresponding to $m$ nodes, the value will be 1 which means $\mathbf{D_S}$ has essentially an identity matrix of size $|M|$ in its top left block. For the $c$ nodes, the value represents the sum of all shares held on all works. A quick check for whether a shares graph is valid would be to check that the partial trace for all $m$ nodes equals the total number of manuscripts $|M|$, and that this equals the trace for the remainder of $\mathbf{D_S}$, i.e. the trace over the $c$ nodes.

The Laplacian matrix $\mathbf{L_S}$ is defined as below where the $M$, $A$, $P$, and $R$ subscripts denote indices corresponding to those nodes respectively.
\begin{equation}
\mathbf{L_S} = \mathbf{D_S} - \mathbf{G_S} = 
\left[
\begin{array}{c|c|c|c}
\mathbf{I_{|M|}} & -\mathbf{S_A} & -\mathbf{S_P} & -\mathbf{S_R}\\
\hline
-\mathbf{S_A} & \mathbf{D_A} & \mathbf{0} & \mathbf{0}\\
\hline
-\mathbf{S_P} & \mathbf{0} & \mathbf{D_P} & \mathbf{0}\\
\hline
-\mathbf{S_R} & \mathbf{0} & \mathbf{0}  & \mathbf{D_R}
\end{array}
\right]
\end{equation}
This matrix is useful for many graph algorithms, with the most notable one being spectral analysis.

\subsection{Spectral Analysis}
\label{ssec:spectral_analysis}
The eigenvalues of the Laplacian matrix $\mathbf{L_S}$ can be used in different ways to ascertain the algebraic connectivity of the graph. Consider the eigenvalue equation for $\mathbf{L_S}$
\begin{equation}
\mathbf{L_S} \mathbf{u} = \lambda_S \mathbf{u},
\end{equation}

\begin{equation}
\mathbf{u}^T \mathbf{L_S}  \mathbf{u} = \lambda_S \, \mathbf{u}^T \mathbf{u},
\end{equation}


Expanding $\mathbf{L}_S$, we have $ \quad
\mathbf{u}^T \mathbf{L_S}  \mathbf{u} = \mathbf{u}^T (\mathbf{D}_S - \mathbf{G}_S) \mathbf{u} 
= \mathbf{u}^T \mathbf{D}_S \mathbf{u} - \mathbf{u}^T \mathbf{G}_S \mathbf{u},
$\\
Then, the eigenvalues $\lambda_s$ is given by \todo{Expand textually, beautify the fraction - AS: Done}
\begin{equation}
\lambda_s = \frac{1}{2 \sum\limits_i u(i)^2}  \sum\limits_{i,j} 
\mathbf{G}_S{ij} \, \big(u(i) - u(j)\big)^2
\end{equation}
Here, $u(i) - u(j)$, represents the distance between nodes $i,j$ in the graph embedding of $\mathbf{G}_S$ based on eigenvector $\mathbf{u}$. 
For a given eigenvalue $\lambda_S$, as we consider nodes $i,j$ with increasing edge weights, $\mathbf{G}_S{ij}$, the nodes are pulled closer together as the distance has to shrink quadratically so that the weighted sum equals $\lambda_S$.

\begin{description}

\item[\textbf{Small eigenvalues ($\lambda_S$).}]
If $\lambda_S$ is small, it means the weighted sum is small for all nodes $i,j$, and hence all nodes are closer together, and the weighted distances between the nodes are all close. Hence, in this case, the eigenvalue represents the number of elements in the sum, and the partitioning chooses the degree of the nodes to cluster, extracting global structure. What this translates to is extracting global communities like academic fields and geographic locations. This could further illuminate collaboration networks, and estimate insularity between fields.

\item[\textbf{Large eigenvalues ($\lambda_S$).}]
If $\lambda_S$ is large, it means the weighted distances between the nodes are larger. Since the distances are weighted by the edge weights, $\mathbf{G}_S$, local structure is emphasized where edge weights determine the partitioning. 
And since for a given $\lambda_S$, the distances shrink quadratically as we increase the edge weights, leading to nodes with higher edge-weights being clustered together. 
For the shares graph, this would mean manuscripts that share high-contribution co-authors would be clustered close together along with their high-contributing co-authors. 
This would end up in a graph split based on contributions to the manuscripts. 
The eigenvector corresponding to the largest eigenvalue would yield a clustering based on the smallest academic entities, such as manuscripts and their highest contributing authors in the Liberata system. Further, the $2,3,\dots$-th largest eigenvalues would correspond to a lab/group, department, institution, etc.

\item[\textbf{Zero eigenvalues and connectivity.}]
The number of $0$ eigenvalues of $\mathbf{L_S}$ represents the number of connected components, i.e., collections of contributors who have interacted with one another either as co-authors, peer reviewers, or replicators. If the second-smallest eigenvalue (the Fiedler value) is non-zero, the graph is connected, and an embedding based on the corresponding eigenvector yields an ordering of nodes reflecting community structure.

\item[\textbf{Multiplicity of zero eigenvalues and domain structure.}]
The multiplicity of $0$ eigenvalues, combined with academic capital, can be used to identify siloed or emerging academic domains. The corresponding eigenvectors reveal fragmented structures and enable further investigation into the associated publications and contributors.

\end{description}


\subsection{Spanning Trees}
\label{ssec:spanning_trees}
\todo{P1: How do people actually measure centrality in graphs - HZ}
Spanning trees have many interesting interpretations and usages in the context of graphs, but one interpretation that is particularly useful in many graph applications and certainly in Liberata's case is a that of connectivity or conductivity. How far a spanning tree can reach starting from any point in the graph provides information about the connectivity of the graph, while the number of possible spanning trees gives information about the number of possible paths, or the degree of connectivity.

For any graph of $n$ nodes, Cayley's formula states that the maximum number of spanning trees in a bipartite graph (i.e. if the graph was a complete bipartite graph) is: 
\begin{equation}
    \tau_c(\mathbf{G_S}) = |M|^{(|C|-1)}|C|^{(|M|-1)}
\end{equation}
with the weighted edge version, if we assume naively equal weights for all edges and still summing to one for every manuscript $m$ being:
\begin{equation}
    \tau_{cw}(\mathbf{G_S}) = s^{(|M|+|C|-1)}|M|^{(|C|-1)}|C|^{(|M|-1)}
\end{equation}

The graph of shares $\mathbf{G_S}$ is expected to be very sparse, and therefore the actual number of tree spanning will be much lower. The actual number of weighted spanning trees, from Kirchhoff's matrix tree formula, is given by
\begin{equation}
    \tau_{kw}(\mathbf{G_S}) = det(\mathbf{L^*_S})
\end{equation}
where $\mathbf{L^*_S}$ is any cofactor of the Laplacian $\mathbf{L_S}$ of the graph $\mathbf{G_S}$, and can be formed by removing any one column and row from $\mathbf{L_S}$. Note that this version takes into the account the strength of connections between nodes. For an unweighted version of the number of spanning trees $\tau_{k}$, set all non-zero values in $\mathbf{G_S}$ to be 1 before computing $\mathbf{L^*_S}$ and then $\tau_{k}$.

Two spanning tree ratios can be then computed. The first is the unweighted ratio of the logarithm of spanning trees.
\begin{equation}
    \text{STR} = \frac{log(\tau_c)}{log(\tau_k)}
\end{equation}
This measure tells us how densely connected (or sparse) the shares graph is compared to a theoretical maximum where everyone collaborated on every paper with everyone else. It can be interpreted as a global collaborativeness coefficient, and can be tracked as a measure of academic trends and research health indicators. Note that the logarithm is used because we expect such extreme sparsity that the measure is only expected to have resolving power when logged.

The second ratio is the weighted ratio.
\begin{equation}
    \text{STR}_w = \frac{log(\tau_{cw})}{log(\tau_{kw})}
\end{equation}
This ratio tells us in relation to the first, the ``conductivity'' or the connectivity density modulated by the strength of connections rather than purely the presence of a connection. If this ratio is much lower than the first, it suggests many uneven share distributions rather than uniform like distributions. Thus the ratio of the two together can be thought of as a measure of asymmetry in share distribution or if taken as analogous to a conductive material, the purity of the conductor.
\begin{equation}
    \text{RSTR} = \frac{STR_w}{STR} \label{eq:relative_spanning_tree_ratio}
\end{equation}

\subsection{Two-Step Shares Graph}
\label{ssec:shares_graph_2_step}
The shares graph $G_S$ shows how people and manuscripts are connected in one step (edge). To observe relationships between contributors, or between manuscripts, or if we want to separate the graph into two connectivity graphs (contributors, and manuscripts), a natural operation is raising $G_S$ to the power of 2. This is called the two-step shares graph, with each element showing the equivalent of random walk likelihood or density of connections from manuscript to manuscript, or contributor to contributor.

\begin{equation}
    \mathbf{G}_s^2 = \mathbf{G}_S^T\mathbf{G}_S = \mathbf{G}_S \mathbf{G}_S  
\end{equation}

The adjacency matrix of the two step shares graph has a block-diagonal structure, with two major blocks on the diagonal corresponding to the two separate subgraphs:
$$\mathbf{G}_s^2 = 
\left[
\begin{array}{c|c}
\mathbf{M}_S^2 & \mathbf{0} \\
\hline
\mathbf{0} & \mathbf{C}_S^2\\
\end{array}
\right]
$$
\begin{itemize}
    \item $\mathbf{M}_S^2$ which has elements representing likelihood of random walking from a manuscript node to a manuscript node along the edges on $G_S$. 
        \subitem The diagonal elements of this component $\mathbf{M}_S^2 (x,x) = \mathbf{M}_S(x,:) \cdot \mathbf{M}_S(x,:)$ represent how concentrated (non-uniform) the shares distribution is for the manuscript $m_x$ for its contributors.
        \subitem The off-diagonal elements $\mathbf{M}_S^2 (x,y) = \mathbf{M}_S(x,:) \cdot \mathbf{M}_S(y,:)$, represent the similarity of contributor shares between manuscripts $m_x$ and $m_y$, with a maximum value (1) attained by an exact match in share split, and a minimum value of (0) attained by no shared contributors. \todo{check this statement for accuracy - AS}
    \item $\mathbf{C}_S^2$ which has elements representing likelihood of random walking from a contributor node to a contributor node along the edges on $G_S$. The $\mathbf{C}_S^2$ is composed of 9 sub-blocks, each with similar but distinct meanings for their values. The interpretations of these block element values are as follows.
    \begin{itemize} 

        \item $\mathbf{S}_A^2$ represents traversal likelihood from an author node to an author node. 
            \subitem The diagonal elements of the block, gives the concentration of distribution of shares by the author across their manuscripts.         
            \todo{These following items need much clearer descriptions of how to interpret the values, including max and min interpretations. - AS}
            $$\mathbf{S}_A^2(x,x) = \mathbf{S}_A(x,:) \cdot \mathbf{S}_A(x,:) , \quad  \forall x \in \{x\in \mathbb{N}, |M| < x \leq |M| + |C|\}$$ 
            The max value here A higher value here means the author's average shares are 
            \subitem The off-diagonal elements of the block,  $\mathbf{S}_A^2(x,y) = \mathbf{S}_A(x,:) \cdot \mathbf{S}_A(y,:) , \quad  \forall |M| < x,y \leq |M| + |C|$, represent similarity of shares split between collaborating authors $c_x$ and $c_y$ across all manuscripts 
        \item $\mathbf{S}_A \mathbf{S}_P$ represents traversing from an author node back to a peer reviewer node. 
            \subitem The diagonal elements of the block, $\mathbf{S}_A \mathbf{S}_P(x,y) = 0 , \quad  \forall |M| < x \leq |M| + |C|, y= |M| + |C| + x$, since this would mean the same contributor was both an author and a peer reviewer for a manuscript. 
            \subitem The off-diagonal elements of the block, $\mathbf{S}_A \mathbf{S}_P(x,y) = \mathbf{S}_A(x,:) \cdot \mathbf{S}_P(y,:) , \quad  \forall |M| < x \leq |M| + |C|, |M| + |C| < y \leq |M| + 2|C|$, represent similarity of shares split between $c_x$ and $c_y$ when $c_x$ is the author and $c_y$ is the peer reviewer, across all manuscripts where that holds true.
        \item $\mathbf{S}_A \mathbf{S}_R$ represents traversing from an author node back to a replicator node. The diagonal and off-diagonal elements hold the same interpretations as for the $\mathbf{S}_A \mathbf{S}_P$ block
    \end{itemize}
\end{itemize} 

\[
\mathbf{G}_S^2 =
\left[
\begin{array}{c|c|c|c}
\mathbf{M}_S^2 & \mathbf{0} & \mathbf{0} & \mathbf{0} \\
\hline
\mathbf{0} & \mathbf{S}_A^2 & \mathbf{S}_A \mathbf{S}_P & \mathbf{S}_A \mathbf{S}_R \\
\hline
\mathbf{0} & \mathbf{S}_A \mathbf{S}_P & \mathbf{S}_P^2 & \mathbf{S}_P \mathbf{S}_R \\
\hline
\mathbf{0} & \mathbf{S}_A \mathbf{S}_R & \mathbf{S}_P \mathbf{S}_R & \mathbf{S}_R^2
\end{array}
\right]
\]

The concentration and similarities of shares distributions computed for contributors in different roles and manuscripts would serve as building blocks for metrics elaborated in \cref{sec:dist_metrics}.  
Further, spectral analysis on the $\mathbf{G}_S^2$ graph, would help in looking at the spectral properties of connectivities between contributors only, and manuscripts only.

\subsection{Condensed Form}
\label{ssec:condensed_form}
The full adjacency matrix $\mathbf{G_S}$ contains a significant amount of structural redundancy due to its bipartite construction. In particular, 10 of the 16 blocks are identically $\mathbf{0}$, and the remaining nonzero blocks consist of three contributor--manuscript matrices and their transposes.

A condensed representation of the shares graph can therefore be defined by retaining only the 3 top-right blocks (out of 16) of $\mathbf{G_S}$:
\begin{equation}
\mathbf{G'_S} =
\begin{bmatrix}
\mathbf{S_A} & \mathbf{S_P} & \mathbf{S_R}
\end{bmatrix}
\label{condensedShares}
\end{equation}

where:
\begin{itemize}
    \item $\mathbf{S_A}$ is the matrix block that encodes shares held by author nodes,
    \item $\mathbf{S_P}$ is the matrix block that encodes shares held by peer reviewer nodes,
    \item $\mathbf{S_R}$ is the matrix block that encodes shares held by replicator nodes.
\end{itemize}

This condensed form captures all edge information in $G_S$, since the remaining nonzero blocks of $\mathbf{G_S}$ are given by the transposes $\mathbf{S_A}^T$, $\mathbf{S_P}^T$, and $\mathbf{S_R}^T$. The full adjacency matrix can therefore be reconstructed as:
\begin{equation}
\mathbf{G_S} =
\begin{bmatrix}
\mathbf{0} & \mathbf{G'_S} \\
(\mathbf{G'_S})^T & \mathbf{0}
\end{bmatrix}
\label{reconstructedShares}
\end{equation}

The advantage of this representation is computational. Storing only $\mathbf{G'_S}$ reduces memory usage and simplifies matrix operations, while preserving the ability to recover the full graph when required for spectral or other analyses introduced in the above subsections.


\newpage
\section{References Graph}
\todo{P4: Read through for coherence, notation consistency, correctness - HZ}
\label{sec:references_graph}
The references graph ${G}_W$ is composed of manuscripts $M$ as nodes, and citations as edges. The edges are denoted by $U$ if they are unweighted, and $W$ if they are weighted. In the traditional unweighted case, there would be a directed edge $u$ with weight $1$ going from manuscript node $m_y$ to $m_x$ if $m_y$ cites $m_x$, while in the weighted case, the edge will have a weight $w_{x,y} \in \mathbb{R}$. Formally, $G_W = (M,W)$ This would normally yield a directed acyclic graph (Directed Acyclic Graph) (assuming no revisions are made). In Liberata, all revisions will be version controlled and citable. 


\todo{Done: Trim it down based on what goes into Section 2 subsections - AS}
As mentioned in \cref{sssec:liberata_weighted_citations}, in the Liberata system, each paper could only could give out a single unit of citation or academic capital (\AC).
This upperbounds the total citation units disbursed by a manuscript to $1$, while in the unweighted case, the total citation units disbursed by a manuscript is equal to the count of all the references $r$. More advanced methods of weighting have been explored in \ref{sec:exploits_&_modifications}.

\subsection{Matrix Representation}
The references graph $G_R$ can also be represented as a square adjacency matrix $\mathbf{G}_{R} \in \mathbb{R}^{|M| \times |M|}$, 
and each element $\mathbf{G}_{W}(x,y)$ will be given by 
\begin{equation}
    \mathbf{G}_{W}(x,y) = \begin{cases}
    w_{x,y} & \text{if }  m_y \text{ cites } m_x\\
    0 & \text{otherwise} \\
    \end{cases} \label{eq:ref_graph_elemwise}
\end{equation} 
Here, $m_x, m_y$ are manuscripts and $w_{x,y}$ is the weighted citation received by $m_x$ from its reference by $m_y$. This matrix will be an upper triangular matrix in the Frobenius normal form as shown below \cite{}. In the simplest case of weighting with the inverse of citations as given in \cref{citationsWeighted}, the adjacency matrix becomes column-stochastic as each column sums to 1.

\todo{Done: Check the representations. Just make $w_{i,j}$ as weights. Specify that in the simplest weighting, $w_{i,j} = \frac{1}{\mathrm{ref}_j}$ -  AS}

\[
\mathbf{G}_W =
\left[
\begin{array}{c|c|c|c|c|c|c}        
0 & w_{1,2} & \cdots & w_{1,y} & \cdots & w_{1,n-1} & w_{1,n} \\
\hline
0 & 0 & \cdots & w_{2,y} & \cdots & w_{2,n-1} & w_{2,n} \\
\hline
\vdots & \vdots & \ddots & \vdots & \ddots & \vdots & \vdots \\
\hline
0 & 0 & \cdots & w_{x,y} & \cdots & w_{x,n-1} & w_{i,n} \\
\hline
\vdots & \vdots & \ddots & \vdots & \ddots & \vdots & \vdots \\
\hline
0 & 0 & \cdots & 0 & \cdots & 0 & w_{n-1,n} \\
\hline
0 & 0 & \cdots & 0 & \cdots & 0 & 0
\end{array}
\right]
\]

\todo{Done: Mention column stochasticity here instead of in fetch vectors -  AS}

The Frobenius normal form would induce a temporal ordering along the rows and columns of the matrix. In this form, the DAG would be an upper triangular matrix with zero along the diagonals, since a manuscript cannot cite itself, and each manuscript may only be cited by other manuscripts published later than itself.
Interesting to note, a lot of the bibliometry-based citation weights, e.g. co-citations or co-references could still be computed from two-step graphs (refer \cref{ssec:gram_matrix} and \cref{ssec:transpose_gram_matrix}) when using the matrix representation.

\subsection{Fetch Vectors}
The value of the edge $w_{x,y}$ induces interesting properties when unit vectors $\mathbf{v}$ are multiplied onto the $\mathbf{G}_W$. Since each column contains the relative references of a single manuscript, the column sum will always be 1. $(\mathbf{G}_W \mathbf{v}_x)\cdot \mathbb{1}^{|M|} = \sum_x w_{x,y} = 1$. This serves as a quick integrity check of the references graph in case there are faulty or misspelled references. 

A left multiplication with the fetch vector $(\mathbf{v}_y\mathbf{G}_W )$ results in a vector with components representing academic capital earned by $y$ from every other work, $\ACEQ_y$. When weighted using \cref{citationsWeighted}, these components represent a distribution of academic capital earned by manuscript $y$ across the works that cite $m_y$ and the component sum yields the total academic capital value accrued by the manuscript $m_y$, $\ACEQ_{y}$ as shown in \cref{eq:ACEQ_1manu_vec}.

\todo{Done: Write that this is the distribution across citing works. You get total AC by element sum of fetch vector. Dont use $1^M$ notation - just use a sum -  AS}
\begin{equation}
    \ACEQ_{y} = \sum (\mathbf{v}_y  \mathbf{G}_W)  \label{eq:ACEQ_1manu_vec}
\end{equation}

Furthermore, since each fetch vector now yields a continuous set of numbers, it would be possible to make observations on more fine-grained assessments of the impactfulness of a manuscript by examining the distribution of impact of that manuscript on other manuscripts. 

\subsection{Compositions of Fetch Vectors} 
Since the fetch vectors represent the accrual of academic capital units by each manuscript, and all the fetch vectors have dimensionality $|M|$, this allows us to collect certain fetch vectors of interest to form certain communities. These collections of fetch vectors would form subspaces within the space of all manuscripts. These collections could be defined by authors, academic fields of research, institutions, research groups, geographic locations, etc. Summary statistics on each collection would give useful information about the interchanges of academic value within each such collection. 

\todo{P3: Too abstract. Change this to be coherent with rest of the sections. Either specify w/ eqns how you would compute, or remove it. - AS}
The partitioning of the reference graph into distinct communities can allow computations of similar measures on overall communities, since each community can also have citation edges coming from and going to other communities. The in-degrees and out-degrees of such communities will allow measures of collaboration and global impact of specific academic communities. These measures can also be spread out over time to analyze the dynamics of academic interest and progress. In a similar vein, within the communities, connectivity measures would elucidate within community collaborations. 


\subsection{Community Detection}
\todo{DONE: P1: Change it to spectral analysis with symmetrized adjacency - AS}
In order to partition the citation graph into different communities, community detection needs to be performed. However, citation networks are DAGs, where the eigenvalues of the adjacency matrix are all zeros, the eigenvectors belong to the null space of the adjacency matrix and can take complex values. 
Therefore, in order to detect citation based communities, we make the assumption that the directed relationship is not important for detecting communities within academic works. Following this, we symmetrize the references graph, and then use the spectral analysis detailed in \cref{ssec:spectral_analysis}.
$$G_W = G_W + G_W^T$$
Since the edges of the references graph are weighted (by default) by the amount of references in each paper, this makes for a natural measure of how strongly the papers are connected to one another, with the strength diminishing as the citing paper cites more papers. Then, the zero, small and large eigenvalues have a similar meaning as in \cref{ssec:spectral_analysis}

\subsection{Centrality Measure}
Centrality measures identify whether a node is crucial in connecting different parts of the graph. 
For references graph, a central node would be the manuscripts that synthesize many past works and add value to conceptual understanding for the newer works.
These manuscripts would represent highly influential works in a field, or influentual interdisciplinary works across fields.
We show how betweenness centrality can be computed here, but other centrality measures, like degree centrality, closeness centrality, or flow-based centrality could also be used in detecting the central nodes.
Betweenness centrality for the references graph can be computed as the ratio of shortest directed paths ($SP$) between 2 manuscripts $m_x, m_z$ which pass through manuscript $m_y$

\begin{equation}
\label{eq:centrality_references}
    \text{Centrality}_y^{\texttt{btw}} = \sum_{\substack{y \neq x \neq z}} \frac{SP_{x,z}(y)}{SP_{x,z}}
\end{equation}

\subsection{Basic Distribution Metrics}
Using the notion of academic capital and relative citation introduced in \cref{ssec:shares_system} allows computation of distribution of capital accrual over the space of manuscripts. 
These metrics can either be computed for individual manuscript $M_j$ cited by manuscripts $\{M_i\}$ to extract the temporal distribution of capital accrual, or extract distributions of capital accrual across different academic communities. 
  
The temporal distribution would quantify test-of-time relevance of the manuscript, while distributions across different academic communities would quantify the breadth of impact of the manuscript (referred to as spatial distribution henceforth, since this would be a span in the space of manuscripts).

Further, distributions could also be computed for different academic communities by using aggregates of fetch vectors. Temporal distributions of capital accrual into academic communities could enable detecting overall interest in the specific academic community, and help formulate birth-death dynamics on communities. The spatial distribution of capital accrual by a community would inform about the breadth of impact of a community as a whole. 


\todo{Done: Move to references graph - AS. Talk more about applications and not just the abstract concept. This feels weak. It is a good thing to note, for the later sections}
\subsection{Powers of References Graph}
\label{ssec:composition_references_graph}
Compositions of $G_W$ can include the two-step graph (n-step graph, in general), and gram matrix computation. The two-steps adjacency of the weighted graph $\mathbf{G}_W^2 (x,y)$ captures the amount of academic capital manuscript $m_y$ contributes  to manuscript $m_x$ through 2-steps connections. The row sums would give the impact of manuscript $m_x$ on manuscripts two generations later. Similarly, $n$-th power of $G_W$, $\mathbf{G}_W^n (x,y)$ reports the impact of manuscript $m_x$ on $m_y$ $n$ steps/generations later. This allows to evaluate long-range academic impact of manuscripts. When using weighted edges with $w_{x,y} \in [0,1]$, the influence would diffuse away rapidly, hence persistence of influence over multiple steps of the reference graph would filter out extremely influential works. However, a much more useful operation would be the gram matrices discussed in \cref{ssec:gram_matrix} and \cref{ssec:transpose_gram_matrix}.

\subsection{Gram Matrix of References Graph}
\label{ssec:gram_matrix}
The gram-matrix of the references graph shows interesting properties, even in the case when edges are unweighted. The following properties can be observed from the matrix given by, $\mathbf{G}_U^T \mathbf{G}_U$

\begin{itemize}
    \item when $x=y$, i.e., for diagonal entries, $\mathbf{G}_U^T \mathbf{G}_U(x,x)$ encodes the number of references of the manuscript $m_x$
    \item when $x \neq y$, i.e., for non-diagonal entries,  $\mathbf{G}_U^T \mathbf{G}_U(x,y)$ gives the number of references manuscripts $m_x$ and $m_y$ have in common. This can be used to estimate a form of \textit{bibliometric coupling} \cite{}
\end{itemize} 

Further, the gram-matrix of the weighted reference graph $\mathbf{G}_W^T \mathbf{G}_W$ has following related properties:
\begin{itemize}
    \item when $x=y$, $\mathbf{G}_W^T \mathbf{G}_W (x,x)$ gives the concentration of academic capital outflow, in terms of the citations it contributes, from the manuscript $x$ without considering its influence on other manuscripts
    \item when $x \neq y$, $\mathbf{G}_W^T \mathbf{G}_W (x,y)$ gives the concentration of academic capital outflow, in terms of the citations it contributes, from the manuscripts $m_x$ and $m_y$ going to some set of common manuscripts $M$ 
\end{itemize}

\todo{Done: Move to References graph -  As}
\subsection{Transpose Gram Matrix}
\label{ssec:transpose_gram_matrix}
The gram-matrix of the transpose of the references graph shows interesting properties too. If the gram matrix of the original graph described concentration of outflow, the gram matrix of the transpose characterizes that of inflow.
For the unweighted reference graph, the gram matrix given by, $\mathbf{G}_U \mathbf{G}_U^T$, can be observed to have the following properties:
\begin{itemize}
    \item when $x=y$, i.e., for diagonal entries, $\mathbf{G}_U \mathbf{G}_U^T (x,x)$ represents the number of citations of the manuscript $m_x$
    \item when $x \neq y$, i.e., for off-diagonal entries, $\mathbf{G}_U \mathbf{G}_U^T (x,y)$ represents the number of manuscripts that cite both $m_X$ and $m_y$
\end{itemize}

Similarly, for the weighted reference graph, the gram matrix given by, $\mathbf{G}_W \mathbb{G}_W^T$, can be observed to exhibit the following properties:
\begin{itemize}
    \item when $x=y$, i.e., for diagonal entries, $\mathbf{G}_W \mathbf{G}_W^T(x,x)$ is proportional to the concentration of academic impact influenced by manuscript $m_x$ independently. When expressed as a fraction of the total academic capital accrued by the manuscript, this represents what percentage of that capital is not shared with other manuscripts.
    \item when $x \neq y$, i.e., for off-diagonal entries, $\mathbf{G}_W \mathbf{G}_W^T(x,y)$ represents the co-concentration of academic impact influenced by manuscripts $m_x$ and $m_y$. It may be better understood as the fraction of the manuscript $m_x$'s academic capital that comes from sources which also cite manuscript $m_y$. 
\end{itemize}

\todo{Read through for coherence, notation consistency, correctness - HZ}
\todo{Done: Change this to Capital Graphs section - AS}
\newpage
\section{Capital Graph}
\label{sec:graph_composition}
Representation of the graphs as matrices allows for various convenient algebraic operations, not only with the same graph, but also between the reference and shares graph. 
To simplify the language and operations in this section, we will refer to graphs $G$, and their adjacency matrix representations $\mathbf{G}$ interchangeably. 
We will proceed to demonstrate compositions of the graphs above and how such compositions can be interpreted.

\subsection{Capital Graph Computation}
\label{ssec:capital_graph}
The Capital graph $G_\ACEQ$ would have a structure similar to the shares graph $G_S$, but with the difference that edge weights represent academic capital received from the manuscripts. 
The capital graph in its condensed form $G_{\ACEQ}'$ can be derived from the condensed shares graph ($\mathbf{G}_S'$) and  references graph ($\mathbf{G}_W$) adjacency matrices using definitions given in \cref{eq:ACEQ_1manu_vec} and \cref{ACDefinition}. 
Formally, we first compute the \AC of all the manuscripts $\ACEQ_M$ using \cref{eq:ACEQ_1manu_vec} by right multiplying it with a vector of $1$s, $\mathbf{1}_{|M|} = (1, \ldots, 1) \in \mathbb{R}^{|M|}$,
Then, we element-wise multiply the vector with each column of $G_S'$, i.e., compute the Hadamard product (denoted $\odot$)   

\begin{gather}
    \ACEQ_M = \mathbf{G}_W \mathbf{1}_{|M|} \\
    \mathbf{G}_{\ACEQ}' = \ACEQ_M \odot G_S' \label{eq:AC_matrix}
\end{gather}

\todo{Done: Make the mathbb 1 more concrete - if it is 1D or 2D. Make sure this works with corrections in the exploits. Separate out the $G_W$ 1 computation as AC. Mentions $Gs_hat$ in Shares graph. -  AS}


\todo{Done: Move the reduced form and reconstruction to Shares graph - HZ}
The full square capital graph matrix $\mathbf{G}_\ACEQ$ can be constructed from its reduced representation $\mathbf{G}_\ACEQ'$ in a similar way as the full square shares graph $\mathbf{G}_S$ would be constructed. Similar to the shares graph, the cumulative capital owned across all manuscripts can be obtained by a sum along columns of this matrix. (Substitute $C \equiv c$ in \cref{ACDefinition}). 

\subsection{Similarities to Shares Graph}
The full Capital graph $G_{\ACEQ}$ will exhibit the following similarities with the full Shares graph $G_S$ 
\begin{itemize}
    
    \item
    \textbf{Adjacency Structure :} The adjacency matrix of the capital graph $\mathbf{G}_{\ACEQ}$ will have similar structure as that of the shares graph in \cref{eq:shares_matrix_form}, however each of the edges will represent the capital value contributor $c$ owns on manuscript $m$, $\ACEQ_{m,c}$.

    \item
    \textbf{Fetch vectors :} Fetch vectors give unnormalized distributions as described in \cref{ssec:fetch_vectors}, with the difference being here that the distributions would be of \AC. However, the row and column sums would not add up to $1$, since the amount of academic capital accrued by each individual manuscript can vary. 

    \item
    \textbf{Fetch vector compositions :} Similar to the description in \cref{ssec:compositions_fetch_vectors}, composition of fetch vectors on the capital graph represent different portfolios (check \cref{ssec:academic_capital_portfolio}).

    \item 
    \textbf{Degree Matrix :} The degree matrix has a similar structure as well, but instead of having cumulative shares, the diagonal elements have cumulative capital, for both manuscripts and contributors.

    \item 
    \textbf{Laplacian and Spectral Analysis :}  Spectral analysis can be performed on $\mathbf{G}_{\ACEQ}$ as well, by normalizing the Laplacian, and using the analysis in \cref{ssec:spectral_analysis}. The only difference from \cref{ssec:spectral_analysis} is that the connectivity structure now is determined in terms of academic capital $\ACEQ$. 

    \item 
    \textbf{Spanning Trees Ratio} Spanning trees ratio can also be computed based on $\ACEQ$ based weightings using \cref{eq:relative_spanning_tree_ratio}.
    
\end{itemize}
\todo{Done: List all similarities at one place  - AS}

\subsection{Two-Step Capital Graph}
The capital graph $G_{\ACEQ}$ can be further composed with itself to form the two-step capital graph, which has a similar node traversal information as the two-step shares graph, but with the edges weighted by academic capital accrued by the contributors on the manuscripts.

The adjacency matrix is equivalent to the gram-matrix, given by, $\mathbf{G}_{\ACEQ}^2 = \mathbf{G}_{\ACEQ}^T \mathbf{G}_{\ACEQ}$, which will have a block diagonal structure, with two major blocks on the diagonal. While the $\mathbf{M}_{\ACEQ}^2$ block will represent the distribution of capital over the manuscripts, similar to the $\mathbf{M}_S^2$ block in \cref{ssec:shares_graph_2_step}, the other diagonal block reveals some interesting quantities. 
\[
\mathbf{G}_{\ACEQ} ^2 =
\left[
\begin{array}{c|c|c|c}
\mathbf{M}_{\ACEQ}^2 & \mathbf{0} & \mathbf{0} & \mathbf{0} \\
\hline
\mathbf{0} & \mathbf{\ACEQ}_A^2 & \mathbf{\ACEQ}_A \mathbf{\ACEQ}_P & \mathbf{\ACEQ}_A \mathbf{\ACEQ}_R \\
\hline
\mathbf{0} & \mathbf{\ACEQ}_A \mathbf{\ACEQ}_P & \mathbf{\ACEQ}_P^2 & \mathbf{\ACEQ}_P \mathbf{\ACEQ}_R \\
\hline
\mathbf{0} & \mathbf{\ACEQ}_A \mathbf{\ACEQ}_R & \mathbf{\ACEQ}_P \mathbf{\ACEQ}_R & \mathbf{\ACEQ}_R^2
\end{array}
\right]
\]
The $\mathbf{C}_{\ACEQ}^2$ block would be composed of 9 blocks again, representing all combinations of starting and ending node categories. We will walk through the first three blocks for traversing from an author node, and remark the values.
\begin{itemize}
    \item $\mathbf{\ACEQ}_A^2$ represents traversing to an author node from another author node. 
    \subitem When both nodes are the same, it represents the concentration of capital accrued on manuscripts. If the author has accrued a lot of capital on all of their manuscripts, the value would be high. This would be a measure of lifetime academic productivity of the contributor as an author. 
    \subitem When the source and destination nodes are different, the number represents the academic value of collaboration between the concerned authors. A higher value indicates that manuscripts involving both the contributors as authors has generated large academic capital, and hence higher impact.
    \item $\mathbf{\ACEQ}_A \mathbf{\ACEQ}_P$ represents traversing to a peer reviewer node from an author node.
    \subitem The diagonal elements would be zero. (Check \cref{ssec:shares_graph_2_step})
    \subitem The off-diagonal elements represent the academic capital produced the contributor interacted as a peer reviewer with the author. Since we expect the cost of peer reviews to be small compared to author shares, a high value here would usually be indicative of collusion or other anomaly, i.e., a high-risk author. This metric would need to be paired with other metrics, i.e., portfolio risk to make a conclusive decision. In the case that collusion is ascertained, the value would represent the academic cost of collusion (in terms of \AC).
    \item $\mathbf{\ACEQ}_A \mathbf{\ACEQ}_R$ represents traversing to a replicator node from an author node. The diagonal and off-diagonal elements have similar interpretations as for the $\mathbf{\ACEQ}_A \mathbf{\ACEQ}_P$ block.
\end{itemize}

\newpage
\section{Portfolio Metrics}
\label{sec:portfolio_metrics}
Portfolios are sets of manuscripts that belong to entities like individual researchers, labs, institutions, geographic regions, time periods, academic fields, etc. Grouping manuscripts into these meaningful sets allows Liberata to produce insightful portfolio level metrics.
\subsection{Academic Capital}
\label{ssec:academic_capital_portfolio}
Recall from \cref{ssec:shares_system} that academic capital is defined as the sum of the product of shares and weighted citations over some set of papers (\cref{eq:AC_Defn}). 
$$\ACEQ_{M,C} = \sum_{m\in M, c \in C} s_{m,c}\cdot w_m$$

\todo{Done - P3: Remove redundancy with section 2 - HZ}
By picking shares belong to different sets of $M$ and $C$, we can measure the following, which are meant to be illustrative and not exhaustive.
\begin{itemize}
    \item\leavevmode Set $M$ to be the papers an individual $C_i=a_i\cup p_i \cup r_i$ has shares on. $\ACEQ \space$ in this case represents the individual's career academic contributions and can be a higher resolution, more accurate metric compared to citation count, H-index, and its derivatives. If $C$ is restricted to just the author node, peer reviewer node, or replicator node of this individual, and one can see the individual's relative contributions in original research, academic review, or replication. Restrict $M$ by papers belonging to academic disciplines, and one can give an accurate breakdown of contributions of the individual to different fields of science.
    
    \item\leavevmode Set $M$ to be the papers that a lab or institution is involved in and $C$ to be the three roles of all members of that institution. $\ACEQ \space$ in this case measures the total contributions of those institutions to academic research, and this quantity can be further broken down by particular labs, by function (authorship, review, replication), and by academic discipline. One could also measure average contribution or collaborativeness of labs or institutions by examining ratios of $\ACEQ_{M,C} \space$ with the $C_{institution}$ contributors to the $\ACEQ_M \space$ for all contributors (i.e. setting $s_{m,c}$ to 1).
    
    \item\leavevmode Set $M$ to be the papers for a particular academic discipline, and one can calculate the relative impacts and sizes of different academic fields.
    
    \item\leavevmode Set $M$ to be papers published within a particular time period, and one can calculate and compare the relative impacts and productivity of different eras within academia. 
    
    \item\leavevmode Set $C$ to be academics from particular geographic regions, and one can calculate the academic output of different parts of the globe, and also further subdivide that by filtering $M$ by in-region and out-region works to see where academic capital is generated and where it flows.
\end{itemize}

In addition to the above, one can imagine mixing intersections of the conditions mentioned above (individuals, institutions, geographic regions, time periods, academic field), and more, to make custom "portfolios" of academic contributions. Formally, a portfolio is a collection of shares on a set of manuscripts $M' \subseteq M$ owned by a set of contributors $C'\subseteq C$.
\begin{equation}
    \label{portfolio_definition}
    \Pi_{M',C'} = \left\{ (m, c) \in M' \times C' \;\middle|\; s(m, c) > 0 \right\}
\end{equation}

\subsection{Academic Taxonomy}
\label{ssec:academic_taxonomy}
The classification of all manuscripts (and portfolios) by academic field implies a taxonomic system to categorize all of academia. Liberata inherits and extends the system of 4 levels of tags from OpenAlex \cite{priem2022openalexfullyopenindexscholarly}, which is called the 4D system, analogous to the GICS system for classifying companies into industries. For the purposes of this paper, the two can be thought of as functionally identical and a sufficiently accurate and detailed classification system for all analysis purposes. The Liberata taxonomy consists of four levels of granularity. 
\begin{table}[h]
    \centering
    \caption{Names, descriptions, and populations of each taxonomic level of Liberata}
    \label{tab:my_label}
    \begin{tabular}{|c|c|c|}
        \hline
        \textbf{Entity} & \textbf{Scope / Granularity} & \textbf{Set Size} \\
        \hline
        Domain: $D_1$ & Analogous to schools or faculties & $\sim 10^0$ \\
        \hline
        Department: $D_2$ & Analogous to institutional departments & $\sim 10^1$\\
        \hline
        Discipline: $D_3$ & Analogous to academic disciplines (majors/minors) & $\sim 10^2$ \\
        \hline
        Direction: $D_4$ & Analogous to research topics/directions & $\sim 10^3$ \\
        \hline
    \end{tabular}
\end{table}

Liberata's tagging system is of a four layer tree structure, where $d_4 \in D_4$ tags are child nodes of $d_3 \in D_3$ tags which are in turn child nodes of $d_2 \in D_2$ which are finally child nodes of $d_1 \in D_1$ tags. No child node can have multiple parent nodes. All manuscripts in Liberata are labeled with one $d_4$ tag upon uploading, and automatically inherit $d_3, d_2, d_1$ parent tags. ``Academic field" will be used in this paper to refer to the set of taxonomic tags associated with a manuscript or person.

\begin{figure}[H]
    \centering
    \noindent
    \begin{minipage}[t]{0.48\textwidth}
    \begin{flushleft}
    \begin{tikzpicture}[
        grow=right,
        level 1/.style={level distance=12mm, ...},
        level 2/.style={level distance=16mm, ...},
        level 3/.style={level distance=20mm, ...},
        level 1/.style={
            sibling distance=80mm,
            every node/.append style={draw=red}
        },
        level 2/.style={
            sibling distance=40mm,
            every node/.append style={draw=green}
        },
        level 3/.style={
            sibling distance=20mm,
            every node/.append style={draw=blue}
        },
        edge from parent/.style={draw, -latex},
        every node/.style={
            font=\small,
            align=center,
            text width=2.4cm,
            draw=black,
            rounded corners
        },
        level 0/.style={
            every node/.append style={draw=black, thick}
        }
    ]
    \node {Engineering}
      child { node {Electrical\\Engineering}
        child { node {Signal\\Processing}
          child { node {Wavelet\\Transforms} }
          child { node {Filter\\Design} }
        }
        child { node {Embedded\\Systems}
          child { node {Microcontrollers} }
          child { node {IoT\\Devices} }
        }
      }
      child { node {Mechanical\\Engineering}
        child { node {Thermodynamics}
          child { node {Heater\\Exchangers} }
          child { node {Entropy} }
        }
        child { node {Fluid\\Dynamics}
          child { node {Turbulence} }
          child { node {Non-Newtonian\\Fluids} }
        }
      };
    \end{tikzpicture}
    \end{flushleft}
    \end{minipage}
    \hfill
    \begin{minipage}[t]{0.48\textwidth}
    \begin{flushright}
    \begin{tikzpicture}[
        grow=right,
        level 1/.style={level distance=12mm, ...},
        level 2/.style={level distance=16mm, ...},
        level 3/.style={level distance=20mm, ...},
        level 1/.style={
            sibling distance=80mm,
            every node/.append style={draw=red}
        },
        level 2/.style={
            sibling distance=40mm,
            every node/.append style={draw=green}
        },
        level 3/.style={
            sibling distance=20mm,
            every node/.append style={draw=blue}
        },
        edge from parent/.style={draw, -latex},
        every node/.style={
            font=\small,
            align=center,
            text width=2.5cm,
            draw=black,
            rounded corners
        },
        level 0/.style={
            every node/.append style={draw=black, thick}
        }
    ]
    \node {Social\\Sciences}
      child { node {Economics}
        child { node {Macroeconomics}
          child { node {Monetary\\Policy} }
          child { node {Fiscal\\Stimulus} }
        }
        child { node {Behavioral\\Economics}
          child { node {Decision\\Theory} }
          child { node {Nudging} }
        }
      }
      child { node {Sociology}
        child { node {Urban\\Studies}
          child { node {Gentrification} }
          child { node {Housing\\Policy} }
        }
        child { node {Social\\Theory}
          child { node {Critical\\Theory} }
          child { node {Postmodernism} }
        }
      };
    \end{tikzpicture}
    \end{flushright}
    \end{minipage}
    \vspace{1em}
    \caption{Example tag tree structure illustrating domain, department, discipline, and direction levels in the 4D system.}
    \label{fig:4D_example}

\end{figure}
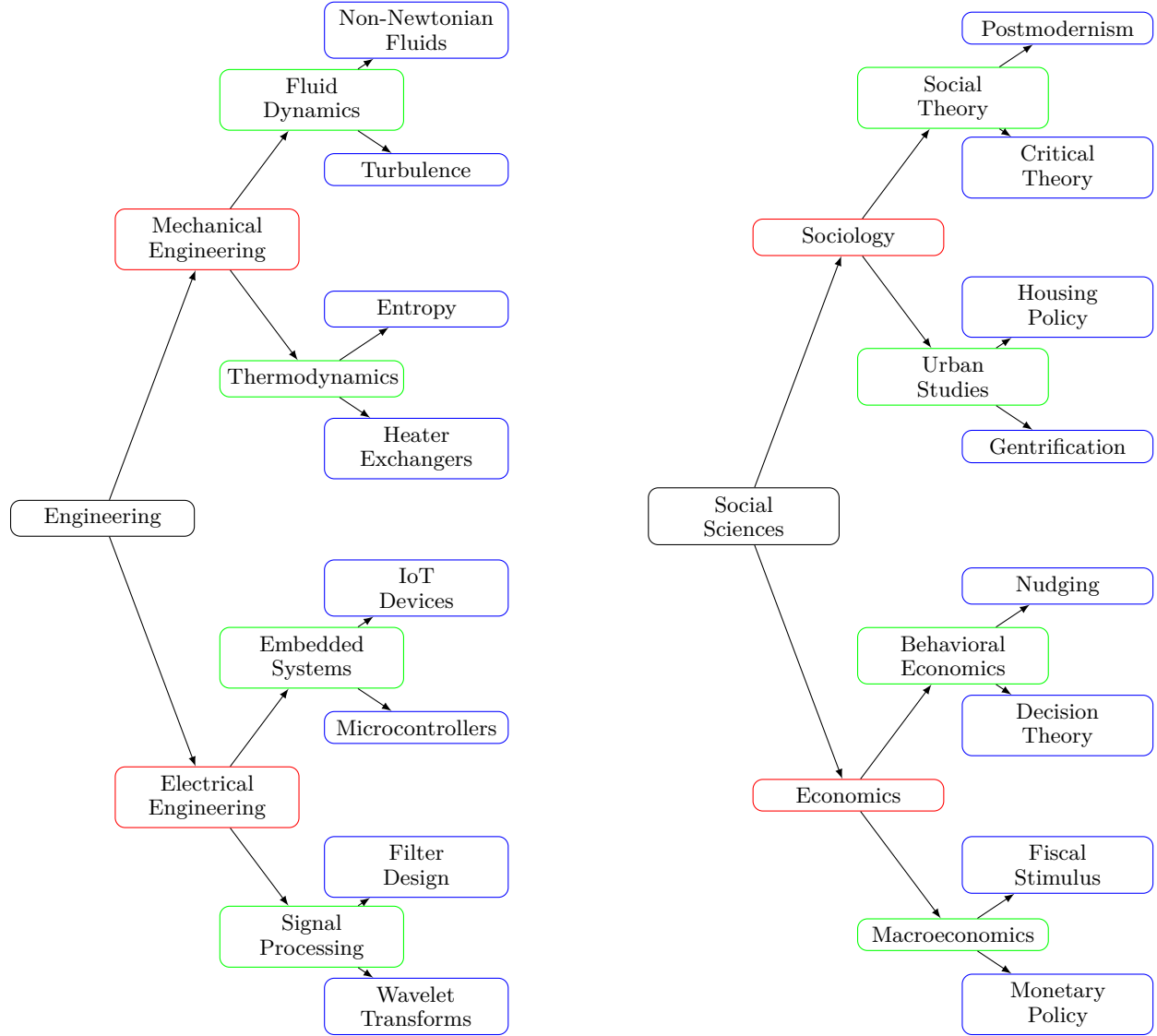

Two examples of the 4D taxonomy are shown in \cref{fig:4D_example}.

\subsection{Portfolio Mix}
\label{ssec:portfolio_mix}
On the Liberata platform, in addition to the 4D tags identifying relevant academic fields, additional tags indicating institution, authors, geographic region, and publication year of each manuscript are used. These tags allow for categorization of different works, and choosing combinations of these tags allows for the construction of an immense range of academic portfolios for further analysis. The first simple analysis is portfolio mix. There are three possible ways to divide an academic portfolio.
\todo{P4: Equation: How would you build a pie chart from this - fetch vectors from Capital graph section - HZ}

\todo{P4:Maybe itemize things - HZ}
Firstly, portfolios can be broken down by academic field to varying levels of granularity according to the 4D system in \cref{ssec:academic_taxonomy}. This is roughly analogous to the multilevel breakdown of equity assets (sector, industry group, industry, sub-industry) by the Global Industry Classification Standard (GICS) widely used by companies such as the S\&P and MSCI. While multiple tags are allowed for each level to facilitate literature searching and browsing, authors will indicate upon uploading which tag at each level is the most relevant to the work, and any academic capital the paper accrues will go to those four tags to prevent multi-counting. 

Secondly, portfolios can be broken down by contributor role. Shares held by any contributor are labeled by whether that contributor is an original author, a peer reviewer, or a replicator of the academic work.

Thirdly, portfolios can be broken down by time period. This breakdown gives a sense of career contributions across time or institution productivity and competitiveness over its existence.

These three breakdowns, field, role, and time, combined with all the tags that can be used to construct portfolios, allow for widely ranging (albeit finite since each set of tags is finite) quantities to be checked and compared.

\todo{P4:Abbreviate - HZ}
\subsection{Relevancy \& Similarity}
\label{ssec:relevancy_similarity}
Given objects categorized by the 4D tag system, a natural way to measure relevance or similarity of works arises. By choosing a tag level of granularity, i.e. $d_4$, one can assemble vectors representing each manuscript by having a 0 for $d_4$ tags that are not tagged onto the manuscript, and a 1 for $d_4$ tags that are tagged onto the manuscript. This can be thought of as a taxonomy vector for that manuscript. By computing the cosine similarity between two taxonomy vectors, one can measure the similarity of the two manuscripts from a value of 0 meaning the two works have no shared tags in common, to a value of 1 meaning the two works have exactly the same set of tags in common.

The choice of granularity by level of tag allows for more fine grained or coarse grained comparisons of manuscripts. The cosine similarity measure can also be extended to portfolios of works by having a taxonomy vector for a portfolio being the weighted average (by academic capital or by shares held) of manuscript taxonomy vectors, allowing for the comparison of similarity between portfolios.

Another method to measure relevancy between two manuscripts $m_i$ and $m_j$, and a way by which to automate the assignment of tags to new works, is to take the rows $i$ and $j$ on the references graph $\mathbf{G}_W$ as co-citation vectors, and compute the cosine similarity between those. If the cosine similarity is found to be close to 1 by some arbitrary threshold, we can assign $m_j$ the same set of tags as $m_i$. A more complicated analysis can also be done looking at the tags of the referenced works of $m_j$, (if $m_j$ is the taxonomically unknown work), and computing a weighted average of all of the manuscripts in $m_j$'s references section and interpreting the resulting value in each component of the vector as the confidence of that tag being appropriate for $m_j$. This can be then presented to $m_j$'s author(s) for verification, or just automatically assigned for backward compatibility or if $m_j$'s authors are unreachable.

The notion of relevancy via cosine similarity, also allows us to construct a relevancy matrix, for search and recommendation purposes. Let $\mathbf{G_D}$ be the relevancy matrix. Then the elements of $\mathbf{G_D}$, $g_{ij}$ are defined as the cosine similarity of the taxonomy vector $\mathbf{d}_i$ of $m_i$ and the taxonomy vector $\mathbf{d}_j$ of $m_j$:

\begin{equation}
g_{ij}  = \frac{\mathbf{d}_i \cdot \mathbf{d}_j}{||\mathbf{d}_i||||\mathbf{d}_j||}    
\end{equation}

Note that by construction, this matrix is symmetric about the main diagonal, and the values along the main diagonal are 1's. By using fetch vectors to retrieve a row $j$, and sorting the indices by relevance, we can find the most relevant works to recommend to the author should they want relevant reading. This measure of relevance can be swapped out for a co-citation version of relevance, instead of the tag based measure of relevance.

\subsection{Expected Returns}
\label{ssec:expected_returns_volatility}
Like with financial assets, the most commonly used portfolios would be that for an individual or institution, although everything mentioned in this subsection does have an analogy to all other possible portfolios or cross sections of manuscripts and their academic capital.

Given a portfolio $\Pi$, one can estimate the expected return or growth in academic capital over time, due to different natural publication rates in different academic fields. In addition, one can also estimate the risk or standard deviation from that mean return. This leads to a relatively simple set of metrics that can help institutions or government bodies gain insight into which academic fields are more impactful per unit risk, informing investment choices when resources are limited. 

The formal definitions and names of these portfolio metrics are:
\begin{equation}
    \label{eq:portfolio_return}
    \text{Returns} = \ACEQ'_{\Pi} = \frac{\Delta\ACEQ_{\Pi}}{\Delta t}
\end{equation}
Where $t$ is selected to be some time period, by default a year. $\ACEQ'$ is a random variable as manuscripts will have some distribution of returns and the exact return for a given manuscript cannot be known a priori.
\begin{equation}
    \label{eq:portfolio_mean}
    \text{Expected return} = \mu_{\Pi} = \mathbb{E}[\ACEQ'_{\Pi}] = \int_{-\infty}^{\infty} \ACEQ'_{\Pi} \cdot f_{\ACEQ'}(\ACEQ'_{\Pi})\, d\ACEQ_{\Pi}'
\end{equation} 
Where $f_{\ACEQ'}(\ACEQ')$ is the probability density function of $\ACEQ'$, not known a priori and gathered from weighted citation (\cref{ssec:shares_system}) and capital graph (\cref{ssec:capital_graph}) data. 

\subsection{Volatility and Risk Asymmetry}
Given the definition for portfolio returns above, the volatility of returns can also be quantified for a portfolio or individual manuscript.
\begin{equation}
    \label{eq:portfolio_volatility}
    \text{Volatility} = \sigma_{\Pi} = \sqrt{\mathbb{E}[\ACEQ'_{\Pi} - \mu_{\Pi}]^2}
\end{equation} 
Volatility or standard deviation, measures how much on average returns differ from the mean return and is a useful risk measure in marketplaces and of portfolio holding entities.

In addition, risk asymmetry, or the asymmetry of academic capital returns can also be measured. This measure can reveal for academic fields, what the underlying skew of distribution of returns is, and whether outliers (fat tail effects) are common or rare.
\begin{equation}
    \label{eq:risk_asymmetry}
    \text{Risk Asymmetry} = \gamma_{\Pi} = \frac{\mathbb{E}[\ACEQ'_{\Pi} - \mu_{\Pi}]^3}{\sigma_{\Pi}^3}
\end{equation}

\subsection{Sharpe's Ratio and Price to Earnings Ratio}
One useful quantity for valuable assets is the returns as a ratio to the volatility, also known as Sharpe's ratio. \cite{sharpe1966mutual} 
\begin{equation}
    \label{eq:sharpe}
    \text{Sharpe's ratio} = \frac{\mu_\Pi}{\sigma_\Pi}
\end{equation}
In Liberata, because all academic capital is generated from manuscripts, and there is no other source of inflation or academic capital generation, the risk free return is 0, which leads to a simplified version Sharpe's ratio.

A natural analogy to the price to earnings ratio for capital assets in finance is the academic capital to returns ratio, which we invert to academic returns to capital ratio (ARC) for more intuitive iterpretability.
\begin{equation}
    \label{eq:ARC}
    \text{ARC} = \frac{\mu_\Pi}{\ACEQ_\Pi}
\end{equation}
This measure gives a simple and intuitive way to tell if a manuscript is in a growth phase of impact (high ARC), tapering phase (low ARC), or stagnant (0 ARC).

\subsection{Allocation Concentration}
\label{ssec:allocation_concentration}
The first step to calculating the allocation concentration of a portfolio is to define the notion of asset weight within a portfolio. 
\begin{equation}
    \label{eq:allocation_weight_m_AC}
    \text{Allocation weight}_{m \in \Pi} = \omega_m = \frac{\ACEQ_m}{\ACEQ_\Pi} = \frac{\ACEQ_m}{\sum_{m \in \Pi}\ACEQ_m}
\end{equation}
This represents the percentage of each portfolio's academic capital $\ACEQ_\Pi$ that comes from each manuscript $m$. Each $\omega_m$ indicates the impactfulness of each asset within the portfolio.

Multiple ways to measure a portfolio's allocation concentration exist. One conventional measure is the Hefindahl-Hirschman Index (HHI), which seamlessly applies to portfolios in Liberata.
\begin{equation}
    \label{eq:HHI_portfolio}
    \text{HHI}_\Pi = \sum_{m \in \Pi}\omega_m^2
\end{equation}
From inspection, the HHI is bounded between (0,1] monotonically, with 1 being the case where there is only one manuscript's shares in the portfolio, and 0 being the case of infinitely many works with a special case of exact 0 value for a case of an empty portfolio.

Another way to measure portfolio allocation is to examine the GINI coefficient of the portfolio.
\begin{equation}
    \label{eq:gini_portfolio}
    \text{Gini}_\Pi = \frac{1}{2|\Pi|}\sum_{k \in \Pi}\sum_{l \in \Pi}|w_k-w_l|
\end{equation}
The Gini is also bounded by [0,1], where 0 would occur if all manuscripts $m$ had the same academic capital $\ACEQ_m$ and 1 would occur if one manuscript accounted for the entire academic capital of the portfolio. This value varies continuously and monotonically with the inequality of weight distributions in the portfolio. Compared with the HHI, this measure will indicate precisely when there is perfect equality with the closed bound at 0, where the HHI will affected by construction by the number of assets in the portfolio, and not only their equality. The HHI will also tend to emphasize more the large allocations, rendering diversity of small allocations more invisible in the presence of some large allocations. These small allocation diversity will be more discernible in the Gini.

A third way to measure portfolio allocation concentration is a normalized entropy measure.
\begin{equation}
    \label{eq:normalized_entropy_portfolio}
    \text{Normalized entropy} = H_\Pi = -\frac{1}{log(|\Pi|)}\sum_{m \in \Pi}\omega_m log(\omega_m)
\end{equation}
This measure is also bounded by 1, but has the opposite interpretation, where 0 is the case of one asset having all the academic capital of the portfolio, and 1 is the case of all assets having equal same academic capital. This measure has some interesting properties regarding impact information carried within the portfolio in the unnormalized form, but in the normalized form, is an excellent measure sensitive to all scales of $\omega$.

It is notable that these above constructions are measuring an academic portfolio's `impact' allocation concentration, which is measured by academic capital. If instead one desired to measure effort allocation concentration within the portfolio, swap the definition of allocation weight $\omega_m$ with
\begin{equation}
    \label{eq:allocation_weight_shares}
    \text{Allocation weight (shares)} = \omega_m' = \frac{s_m}{s_\Pi}=\frac{s_m}{\sum_{m \in \Pi}s_m}
\end{equation}
where $s_m$ and $s_\Pi$ are shares held in manuscript $m$ in portfolio $\Pi$ and the latter is the simple sum of all the shares held across all manuscripts in the portfolio (shares are already normalized values that sum to unity on a manuscript). 

In addition, these weights can be recalculated so that they represent not individual papers, but different tags. This would be done by batching manuscripts into primary tag objects ($d_1, d_2, d_3, d_4$) and computing weights by these groups, rather than by $m$.
\begin{equation}
    \label{eq:allocation_weight_dk_AC}
    \text{Allocation weight ($d_k | k \in \{1,2,3,4\}$, $\ACEQ$)} = \omega_{d_k} = \frac{\ACEQ_{d_k}}{\ACEQ_\Pi}=\frac{\ACEQ_{d_k}}{\sum_{d_k \in \Pi}\ACEQ_{d_k}}
\end{equation}
\begin{equation}
    \label{eq:allocation_weight_dk_shares}
    \text{Allocation weight ($d_k | k \in \{1,2,3,4\}$, $s$)} = \omega_{d_k}' = \frac{s_{d_k}}{s_\Pi}=\frac{s_{d_k}}{\sum_{d_k \in \Pi}s_{d_k}}
\end{equation}
Thus it is possible to measure for any portfolio holding entity their concentration of contributions in impact (\cref{eq:allocation_weight_dk_AC}) or effort (\cref{eq:allocation_weight_dk_shares}) in each academic domain $d_1 \in D_1$, department $d_2 \in D_2$, discipline $d_3 \in D_3$, and direction $d_4 \in D_4$.

\subsection{Diversification Ratio}
\label{ssec:diversification_ratio}
The diversification ratio (DR) is a measure of how much diversification benefits a portfolio by comparing the weighted average volatility of the individual assets to the actual risk of the portfolio. One expects that if fields of science are not perfectly correlated in their average returns, that diversification would yield some benefit measurable by this ratio.
\begin{equation}
    \text{Diversification ratio} = \frac{\sum_{m \in \Pi} \omega_m\sigma_m}{\sigma_\Pi}
\end{equation}
This ratio is lowerbounded by 1, which is the case that all manuscripts are perfectly correlated in their returns. This is likely only possible for small portfolios that have works within a very narrow niche, such that all works are likely to be cited by new works in that narrow niche. The larger the DR, the more the diversification from existing assets in the portfolio is benefiting the portfolio in terms of reducing volatility. This measure could be used to quantify the volatility of scientific careers and reveal the effects of risk for different degrees interdisciplinary collaboration. Together with the allocation concentration metrics, it is possible to reveal patterns in scales of success and impact for varying levels of interdisciplinary academic research work.

\subsection{Funding and Time Research Efficiency}
\label{ssec:research_efficiency}
Two useful ways to measure how efficiently research is conducted is to look at the amount of academic capital $\ACEQ$ generated for the time and/or money invested into the research. Funding research efficiency measures how much $\ACEQ$ each dollar (or other currency or choice) of research funding generates on a given portfolio $\Pi$. 
\begin{equation}
    \label{eq:funding_efficiency}
    \text{Funding Efficiency} = \epsilon_{\$,\Pi} = \frac{\ACEQ_{\Pi}}{\$_{\Pi}}
\end{equation}
This can be measured at the global level, which would quantify how much academic impact taxpayer funding is generating across the world, but can also be calculated for more granular portfolios. In cases where an academic work has multiple sources of funding, Liberata will ask the uploading author to define the proportional share of funding of each source contributed to the work to avoid double counting. If this information is not given, Liberata scales the funding tied to each author by their contribution share. If any single author has multiple funds tied to them, Liberata will take the simple average of the funding to figure subdivide the capital attribution to each source of funding.

Time research efficiency measures how much $\ACEQ$ is generated per unit time for a given entity's (i.e. individual, institution, geographic region, academic field, etc.) portfolio $\Pi$.
\begin{equation}
    \label{eq:time_efficiency}
    \text{Time Efficiency} = \epsilon_{t,\Pi} = \frac{\ACEQ_{\Pi}}{\Delta t_{\Pi}}
\end{equation}
For each paper uploaded onto Liberata, the uploading author is asked to indicate when the work began. $\Delta t_{\Pi}$ is the time from the beginning of the earliest work in the portfolio to the upload date of the latest work. 

These metrics are intended to quantify how effectively academic entities can convert time and funding into academic impact, and depending on choice of portfolio, can have different yet meaningful interpretations.

\subsection{Peer Reviewer and Replicator Reliability}
One way of splitting up a portfolio for an individual contributor is by their contribution role (author, peer reviewer, replicator). On Liberata, if any work is retracted, the value of shares on that work automatically go to zero. Liberata discerns the individual's reliability from two metrics. The first is the proportional loss metric, which is defined below.
\begin{equation}
    \label{eq:relability_loss}
    \text{Proportional Loss} = \%_L = \frac{\ACEQ_{lost}}{\ACEQ_{remaining}}
\end{equation}
This quantity for measuring a peer reviewer's reliability would have that individual's total academic capital from peer review lost in the numerator, and total academic capital from peer review remaining in the denominator. Likewise the proportional loss for replicators are computed the same way, just on a replication academic capital. This quantity is meant to capture the tendency to not catch serious errors in reviewed or replicated works relative to tendency to catch them.

The second number is the proportional split metric, which is defined below.
\begin{equation}
    \label{eq:relability_split}
    \text{Proportional Split} = \%_S = \frac{\ACEQ_{task}}{\ACEQ_{total}}
\end{equation}
This quantity for measuring a peer reviewer's reliability would have that individual's academic capital received from peer review in the numerator, and their total academic capital in the denominator. Likewise, replicator reliability is calculated the same way, just with academic capital gained from replication in the numerator. This quantity is meant to capture how frequently a reviewer or replicator is doing original research versus quality control.

These two numbers are not combined but instead both available for view on the Liberata platform for marketplace goers and readers to better judge the reliability of quality control done. In addition, documents on Liberata are version controlled, allowing users to read the peer reviewer comments as well as the manuscript prior and post each revision. Replications are written up and attached to the document under the automatically added replications section. An important difference is that on Liberata, peer reviewers are still anonymous, but replicators are known to authors and readers.

\todo{Done: Normalize by field rate of returns $m_2$. Write ddt assign variables for slopes as ddt and mention how its computed - AS}
\subsection{Impact of Quality Control}
For any manuscript $m$, we can fit curves to the amount of returns $\ACEQ_m'$ it acquires over time $t$. The returns are expected to follow continuous and differentiable trajectories under normal circumstances, but may become non-differentiable due to the impact of a quality control event, among other things. Let's say for a $\Delta t$ interval surrounding around a quality control event at time $t$, the returns changes at $t$ as $\Delta \frac{d}{dt}\ACEQ_{m} (\Delta t)$, and in the same interval, the returns of the field $d_4$ changed by $\Delta \frac{d}{dt}\ACEQ_{d4} (\Delta t)$ then we compute the impact as the change in slope normalized by the change in the slope for the field.
\begin{equation}
    \label{eq:impact_qc}
    \text{IQC} = \frac{\Delta \frac{d}{dt}\ACEQ_{m} (\Delta t)} {\Delta \frac{d}{dt}\ACEQ_{d4} (\Delta t)} = \frac{\frac{d}{dt}\ACEQ_{m}(t + \Delta t/2) - \frac{d}{dt}\ACEQ_{m}(t - \Delta t/2)}{\frac{d}{dt}\ACEQ_{d4}(t + \Delta t/2) - \frac{d}{dt}\ACEQ_{d4}(t - \Delta t/2)}
\end{equation}
Then, if the quality control event (completed reviews or replications) helped in making the case in the article much stronger, then the
IQC would be large. Thus, this value can be used to attribute the impact of the quality control activity to
the QC service providers (i.e., reviewers and replicators). Using the shares owned by the reviewers/replicators on the manuscript $m$, the quantity of impact can also be suitably attributed to the contribution of each reviewer/replicator. 

\subsection{Collections \& Journals}
\label{ssec:collections_and_journals}
Many traditional scientometrics are well defined for only individuals. For example, citation counts and H-index are easy to compute for individuals, but not well defined for institutions, geographic regions, time periods, or academic fields. If a manuscript has two authors from institution A, and one from institution B, it is not clear for citation count or H-index if this manuscript should be counted once or twice for institution A, with both having problematic implications for A's credit relative to B.

In the Liberata system however, the notion of portfolios extend naturally to any set of shares. Portfolios need not be for only individuals. This allows powerful ways of comparing many different types of entities, and also enables a much simpler and consistent way of defining journals. In the Liberata system, journals are defined by collections of $D_4$ tags. Any manuscript with at least those tags automatically is populated into the journal. (Other filter tags exist for peer reviewed, replicated, etc.) For a journal with manuscripts $M_J$ and total academic capital $\ACEQ_J$, the impact factor of a journal is replaced with the average academic capital $\mathbb{E}(m_J)$ of the manuscripts in the journal.

\begin{equation}
    \mathbb{E}(m_J) = \frac{\ACEQ_J}{|M_J|}
\end{equation}

This measure can be used to measure average impactfulness of institutions, geographic regions, academic fields, and intersections of these and more, by arbitrary selection of which manuscripts to include in the collection. All other metrics in this section also are applicable to any collection, which are ultimately just a type of portfolio.

\newpage
\section{Market Metrics}
\label{sec:market_metrics}
\subsection{Fair Market Prices}
\label{ssec:fair_market_price} 
For a transaction to occur on the Liberata platform, both the author and the peer reviewer or replicator must have positive expected value from its execution. As more transactions accumulate on Liberata marketplaces, a more confident value can be assessed for the fair market price (FMP) of peer review and replication services for each academic field. The definitions of the peer review FMP is:
\begin{equation}
    \text{FMP}_{p,d} = \mathbb{E}[s_{p,d}] = \Bar{s}_{p,d}
    \label{eq:fmp_reviews}
\end{equation}
where $s_{p,d}$ denotes the shares that an average peer reviewer $p$ in the academic field $d$ holds on a work within $d$. Likewise, the definition of FMP for replication is:
\begin{equation}
    \text{FMP}_{r,d} = \mathbb{E}[s_{r,d}] = \Bar{s}_{r,d}
    \label{eq:fmp_replications}
\end{equation}
where $s_{r,d}$ denotes the shares that an average replicator $r$ in the academic field $d$ holds on a work within $d$. Note that we expect the two FMP values to be different for any $d$, because in general, peer review and replication take different amounts of effort to do. It is expected that for most fields, a successful replication will take more effort, but also be a much stronger signal for accuracy for any given work. Thus, we expect that $\text{FMP}_{r,d} > \text{FMP}_{p,d}$ in most cases, but this fundamentally does not always have to be true. It would be interesting to examine the fields where the reverse is true, such as a field where one can readily rerun simple code to check the author's results. In such fields, it may be possible to drive $\text{FMP}_p$ to 0 unless the reviewers are able to add value to the work besides accuracy, such as clarity or interpretation nuance. The FMP values capture information about the perceived value add that peer review and replication has for a work by the members of that academic community who participate in the marketplace, as well as the risk of that work.
\todo{P4: Check for narrative cohesion - HZ}

\Cref{ssec:portfolio_mix} also implies that we can have different `depths' (domain, department, discipline, direction) of FMP with different values, for each of the four levels of tags. This provides a useful tradeoff between accuracy of the FMP measure, versus the simplicity of the computation.

\subsection{Risk Premiums}
\label{ssec:risk_premiums}
Risk premium is the difference between the price (in shares) that a particular author or set of authors pay on average and the fair market price of the author(s)' academic field. The definition for risk premium for the peer review marketplace is:
\begin{equation}
    \text{Risk premium}_{p,d}=\psi_{p,d}(\hat{A})=\hat{s}_{p,d}-\Bar{s}_{p,d}
    \label{eq:risk_premium_reviews}
\end{equation}
where $\hat{s}_{p,d}$ denotes the average shares that author(s) $\hat{A}$ pay for peer review $p$ within academic field $d$, and $\Bar{s}_{p,d}$ is the FMP for peer review in that field. Likewise, the definition for risk premium for the replication marketplace is:
\begin{equation}
    \text{Risk premium}_{r,d}=\psi_{r,d}(\hat{A})=\hat{s}_{r,d}-\Bar{s}_{r,d}
    \label{eq:risk_premium_replications}
\end{equation}

This quantity measures how much additional risk the community within $d$ perceive for the author's work above the mean. (Note: this quantity can be negative if the author(s) $\hat{A}$ are seen as less risky than the average member of the field.) Higher risk premiums can indicate that the author(s) may
\todo{P4: Itemize this better - HZ} 
(1.) have a poor track record of accurate work, (2.) are a newcomer to the field, (3.) have trouble clearly writing descriptions of their work, and/or (4.) are just doing work that is perceived to be less impactful. Whatever the reason, and their might be other reasons than the four identified here, peer reviewers and/or replicators would rather do quality control for other works in the field given the same share compensation and the additional incentive needed is useful information. 

This quantity can be ascertained for any group of authors, such as all the researchers of an academic institution. The risk premiums paid by institutions can represent (1.) the track record of accurate work from that institution, (2.) the clarity of writing and presentation of work from that institution, (3.) the ease of working with authors from that institution, and (4.) perceived impactfulness of the work from that institution. This factors are not an exhaustive list but are or are among the expected major contributors to risk premium. Additionally, within an institution, risk can be calculated for different departments by sectioning authors and works according to their $d_2$ primary tags, allowing for the identification of particularly distinguished fields within the institution, i.e. the institution's specialty. 

\todo{P4: Check narrative flow - HZ}
At present, it is very difficult to discern research risk for institutions, as the available numbers of publication rate, journal impact factors, and citation counts, are individually highly exploitable and loosely correlative to their intended quantity of measure, but also have myriad interpretations that are all valid, leading to a high noise to signal ratio for determining academic risk and poor consensus amongst academics of what the story behind each number means. By contrast, the way the Liberata marketplace and metrics are constructed, if risk premium (perceived risk) is actually inaccurate to true risk, there is additional expected academic capital to be arbitraged from doing or avoiding peer review or replication for that undervalued or overvalued work respectively, allowing the system to self correct inaccurate pricing through the classic market forces that lead to price discovery.

\todo{Done:This should be in mechanisms paper - HZ}

\subsection{Relative Performance and Risk Adjusted Performance}
A simple relative performance measure for any manuscript is to take its expected returns for academic capital $\mu_m$ defined in \cref{ssec:expected_returns_volatility} as a ratio of the expected returns $\mu_{d(m)}$ of the average manuscript in $m$'s academic field $d$, as denoted by $m$'s primary domain, department, discipline and direction tags.
\begin{equation}
    \text{Relative performance}_m = \rho_m = \frac{\mu_m}{\mu_{d(m)}}
\end{equation}
This measure quantifies the gains in academic capital per unit time for the manuscript as a ratio to the field manuscripts' average, which is a useful and intuitive way to indicate the quality of a manuscript. By extension, the relative performance of a portfolio $\Pi$ is the weighted sum of the relative performances.
\begin{equation}
    \text{Relative performance}_\Pi = \rho_\Pi = \frac{1}{\sum_{m \in \Pi} s_m} \sum_{m \in \Pi} s_m\frac{\mu_m}{\mu_{d(m)}}
\end{equation}
Where $s_m$ denotes \% shares the portfolio holds in manuscript $m$. The portfolio relative performance is a useful, intuitive way to represent the performance of any entity that can hold a portfolio of shares on manuscripts, such as individuals, institutions, etc.

Additionally, the performance of any portfolio $\Pi$ can be measured in a risk adjusted way by using a redefined capital asset pricing model (CAPM) \cite{sharpe1964capital} where the risk free return is set to 0, equivalent to the assumption that there is no generation of academic capital outside of manuscripts, which is true by construction of the Liberata system. First, we define the risk adjusted excess return $\alpha_m$ for any manuscript. This quantity represents the extra return above expected for the riskiness of the manuscript. 
\begin{equation}
    \alpha_m = \mu_m-\beta_m\mu_{d(m)}
\end{equation}
where $\beta_m$ is the sensitivity of $\mu_m$ and $\mu_{d(m)}$, defined as below. 
\begin{equation}
    \beta_s = \frac{\text{Covariance}(\mu_m,\mu_{d(m)})}{\text{Variance}(\mu_{d(m)})}
\end{equation}
By extension, the excess risk adjusted returns for a portfolio is the weighted sum of the excess risk adjusted returns for each manuscript within that portfolio.
\begin{equation}
    \alpha_\Pi =\sum_{m \in \Pi}s_m\alpha_{m} = \sum_{m \in \Pi} s_m(\mu_m - \beta_m\mu_{d(m)})
\end{equation}
Risk adjusted relative performance then is taken to be for a single manuscript:
\begin{equation}
    \text{Risk Adjusted Relative Performance}_m = \Tilde{\rho}_m = \frac{\alpha_m}{\mu_{d(m)}}
\end{equation}
\todo{P4: More text and better grouping - HZ}
and by extension, for a portfolio:
\begin{equation}
    \text{Risk Adjusted Relative Performance}_\Pi = \Tilde{\rho}_\Pi = \sum_{m \in \Pi}\frac{\alpha_m}{\mu_{d(m)}}
\end{equation}
These metrics are a simple but powerful way for Liberata to quantify the relative performance of any manuscript or portfolio of manuscripts that is field agnostic.


\newpage
\section{Distribution Metrics}
\label{sec:dist_metrics}
The distributions of shares on manuscripts, academic domains, departments, disciplines, and directions, as well as the authors, peer reviewers, replicators, institutions, geographic regions, time periods that hold those shares, make for a rich tapestry of distribution metrics in addition to the ones already mentioned in prior sections.

\subsection{Author Contribution Distributions}
\label{ssec:author_contribution_distributions}
It is expected that different academic fields will have different natural distributions for authors. As more manuscripts accumulate on Liberata, archetype histograms and distributions can be formed for author shares on different academic fields. In addition, because valid distributions can be formed on any set of manuscripts, distributions can be made for any valid portfolio of manuscripts as defined in \cref{sec:portfolio_metrics}. 

With the author distributions, we can calculate the inequality of author share splits using metrics with similar formulations to \cref{ssec:allocation_concentration}.
\todo{P4: consolidate/abbreviate - HZ}
The typical author share distribution by HHI for a given field $d$ is:
\begin{equation}
    \text{HHI author shares}(d)=\frac{1}{|M_d|}\sum_{m \in M_d}\sum_{a \in A_m}s_{a,m}^2
\end{equation}
This quantity can be compared to the HHI for a portfolio or a single manuscript to check for anomalies between an author's portfolio (\cref{eq:HHI_authors_portfolio}) or a share distribution of a single manuscript (\cref{eq:HHI_authors_manuscript}) compared to that author's academic field.

\begin{equation}
    \label{eq:HHI_authors_portfolio}
    \text{HHI author shares}(\Pi)=\frac{1}{|M_\Pi|}\sum_{m \in M_\Pi}\sum_{a \in A_m}s_{a,m}^2
\end{equation}

\begin{equation}
    \label{eq:HHI_authors_manuscript}
    \text{HHI author shares}(m)=\sum_{a \in A_m}s_{a}^2
\end{equation}
Then, the HHI difference (HHIRD) measures discrepancy between the HHI of the field and that of the manuscript or portfolio in a single number.
\begin{equation}
    \label{HHIRD}
    \text{HHID} = |\text{HHI}_\Pi - \text{HHI}_d|
\end{equation}
Similarly to HHI, other distribution comparisons are possible by taking the Gini coefficient in \cref{eq:gini_portfolio} or the entropy measure in \cref{eq:normalized_entropy_portfolio} with allocation weights by shares (\cref{eq:allocation_weight_shares}) for the author shares in a field $d$ and compare it to that of an arbitrary portfolio or manuscript by difference or ratio respectively. These metrics can call be used as an indicator of anomalous author share splits, which can indicate situations including but not limited to a budding scientific field unlike its parent field in work requirements, or a supervisor that is able to leverage power dynamics to gain shares, or paid for credit manuscript farm products, etc.

\todo{Done: Example graph showing what does a sinking, growing and steady field look like. Just artificially make them  -AS}
\subsection{Population pyramids}
\begin{figure*}[t]
    \centering
    \includegraphics[width=0.95\textwidth]{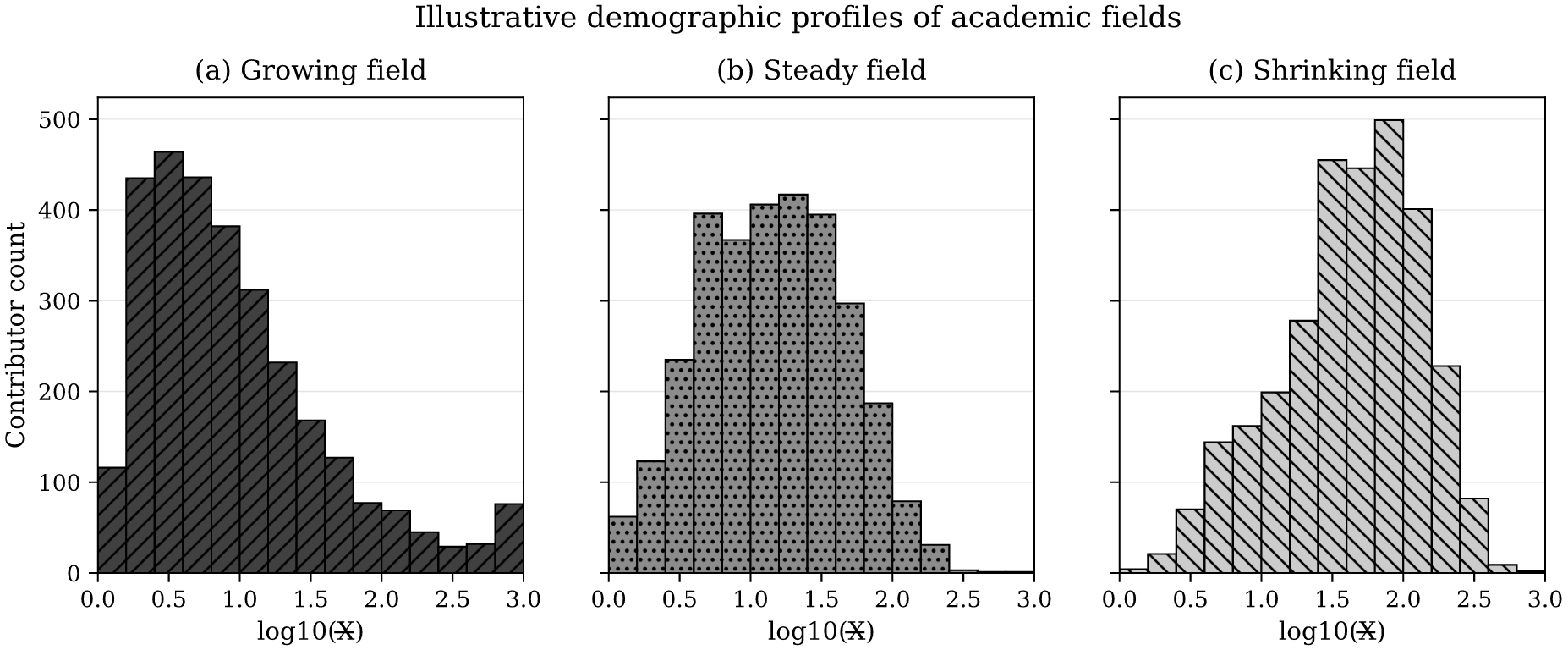}
    \caption{Illustrative demographic profiles of academic fields.}
    \label{fig:field_demography}
\end{figure*}

One way of representing the demographics of academic fields or of academia in its entirety is to make population pyramids or histograms more generally of contributors based on \ACEQ. For these histograms, we choose $log(\ACEQ)$ as the measure axis and contributor count on the frequency axis. This choice reflects that researcher academic capitals like citation counts are likely to be Pareto distributed. \cite{Price1976, Redner1998} These histograms then, with appropriate number of bins, serve as a visualization of whether a field is growing, shrinking, or stagnant (refer \cref{fig:field_demography} for an illustrative of what the histograms might look like). If a field is growing in population, one would expect relatively many young researchers with low $\ACEQ$, whereas with a shrinking field, one expects relatively many older researchers with high $\ACEQ$, analogous to the interpretations for population pyramids in traditional demographics.

\newpage
\section{System Health Metrics}
\todo{P4: This section should come after corrections - AS}
\subsection{Academic Capital Growth Rates}
\todo{P4: This section can then talk about how different corrections can impact capital growth rates -  AS}
As more academic works are produced, the total academic capital in the entire system is expected to grow. The rate of this growth can signal overall system health by measuring academic impact produced at a global level.
This signal becomes more faithful to the actual health of the system as better quality control is implemented.
This can be computed using \cref{eq:portfolio_return}, where the portfolio becomes all the manuscripts that were published a unit time ago, $M' = M_{t-1}$, and $\Delta t =1$. 
By default, $\Delta t$ would be a year, following the convention in \cref{ssec:portfolio_mix}.

\todo{P3: Check for notation, coherence, correctness. Better people are doing QC, or people are getting good at QC}
\subsection{Fair Market Price Shrinkage Rate}
\label{ssec:risk_premium_shrinkage_rate}
The fair market price of peer reviews and replications signal the actual costs associated with reviewing and replicating academic results.
When it becomes easier to derive academic capital from quality control, the prices are expected to see a downward trend until plateauing. 
This metric would measure the rate of negative change in fair market price from \cref{eq:fmp_reviews}, \cref{eq:fmp_replications} for the whole system year-over-year.
Let the change in fair market price of replications over the globe in a unit time period be denoted by $\Delta \Bar{s}_r{(t)} = \Bar{s}_r{(t)} - \Bar{s}_r{(t-1)}$, where $\Bar{s}_r(t) = \sum_{d \in D}  \Bar{s}_{r,d}^{(t)}$, and similarly, the change in fair market price of reviews be denoted as, $\Delta \Bar{s}_p{(t)} = \Bar{s}_p{(t)} - \Bar{s}_p{(t-1)}$. Then, the global shrinkage rates are given by:

\begin{equation}
    \Bar{s}_p' = \frac{- \Delta \Bar{s}_p{(t)}}{\Delta t}
    \label{eq:fmp_reviews_shrinkage}
\end{equation}

\begin{equation}
    \Bar{s}_r' = \frac{- \Delta \Bar{s}_r{(t)}}{\Delta t}
    \label{eq:fmp_replications_shrinkage}
\end{equation}

A weighted global shrinkage rate can be further computed by taking a weighted sum over both $\Bar{s}_p'$ and $\Bar{s}_r'$, weighted by $0<\psi<1$.

\begin{equation}
    \Bar{s}' = \psi \Bar{s}_p' + (1 - \psi) \Bar{s}_r'
    \label{eq:weighted_fmp_shrinkage}
\end{equation}

\subsection{Geographic Capital Distributions}
This would quantify the amount of academic capital per capita, and per GDP produced at different geographic locations. This can also be split by the percentage contribution of academic fields towards the total academic capital of a region. This would inform us about the emergence of specific locations on the globe that would be more specialized, as opposed to more generalists, and if there is a preference for one or the other based on regional resources or economic growth trajectories.

Consider $\Pi_{x}, \ACEQ_{x}$ be the portfolio, and the academic capital of an institute $i$ in the geographic region $\theta$. Then, the total capital of the region is simply a sum over \AC from each institution in the region.
\begin{equation}
    \ACEQ_{\theta} = \sum_x \ACEQ_{x}
\end{equation}
Without loss of generality, consider $\ACEQ_{d}$ be the academic capital from an academic field of depth $d \in D$, where $d$ can be any depth, $d_{1,2,3,4} \in D_{1,2,3,4}$ then,

\begin{itemize}
    \item $\ACEQ_{\theta} = \sum_{D} \ACEQ_{d}$, Total academic capital of a region should be equal to the sum of the academic capital split across all fields at a certain depth
    \item The proportionate contribution of $d$ can be computed as $\ACEQ_{d} / \ACEQ_{\theta}$
    \item The concentration of academic capital contribution of a region can then be computed using Herfindahl-Hirschman index \cite{}
    \begin{equation}
        \text{HHI}_{D} = \sum_{d \in D}  \left( \frac{\ACEQ_{d}}{\ACEQ_{\theta}} \right) ^2
        \label{eq:HHI_regional}
    \end{equation}
\end{itemize}

In addition, consider the per capita academic capital of a region $\theta$ to be 
$$\ACEQ_{\text{PC}, \theta} = \ACEQ_{\theta} / \text{pop}_{\theta}$$ 
Then, the inequality of academic capital contribution across regions can then be computed by the Gini coefficient, 
\todo{P3: Check notation, and correctness - AS}$Gini_{\text{PC'}}(\Theta)$ \cite{Gini1912}, where all regions $\theta \in \Theta$ are ordered in increasing order of $\ACEQ_{\text{PC}, \theta}$.

\todo{Done: Put the fraction as $1/\ACEQ_{\Theta}$ -  AS} 
\todo{P3: Also consolidate the equations and improve narrative flow - AS}
\begin{equation}
    Gini_{\text{PC}}(\Theta) = \frac{1}{\ACEQ_{\Theta}} {\displaystyle \sum_{x=1}^{|\Theta|} \sum_{y=1}^{|\Theta|} abs( \ACEQ_{\theta,y} \cdot \text{pop}_{\theta,y} - \ACEQ_{\theta,x} \cdot \text{pop}_{\theta,x})}
    \label{eq:Gini_regional_percapita}
\end{equation}

 We can also compute the per capita over subset of contributors $\hat{C}_{\theta}$ from the region instead of the whole population, $$\ACEQ_{\text{PC}', \theta} = \ACEQ_{\theta} / |\hat{C}_{\theta}|$$, with Gini coefficient as,
 \begin{equation}
    Gini_{\text{PC'}}(\Theta) = \frac{1}{\ACEQ_{\Theta}}\displaystyle \sum_{x=1}^{|\Theta|} \sum_{y=1}^{|\Theta|} abs( \ACEQ_{\theta,y} \cdot |\hat{C}_{\theta,y}| - \ACEQ_{\theta,x} \cdot |\hat{C}_{\theta,x}|)
    \label{eq:Gini_regional_per_contributor}
\end{equation}

Similarly, the regions $\theta$, can also be arranged in ascending order of per GDP academic capital, $Gini_{\text{GDP}}(\Theta) = \ACEQ_{\theta} / \text{GDP}_{\theta}$. Then the Gini coefficient is given by,
\begin{equation}
    Gini_{\text{GDP}}(\Theta) = \frac{1}{\ACEQ_{\Theta}}\displaystyle \sum_{x=1}^{|\Theta|} \sum_{y=1}^{|\Theta|} abs( \ACEQ_{\theta,y} \cdot \text{GDP}_{\theta,y} - \ACEQ_{\theta,x} \cdot \text{GDP}_{\theta,x})
    \label{eq:Gini_regional_perGDP}
\end{equation}

While the Gini coefficients in \eqref{eq:Gini_regional_perGDP} and \eqref{eq:Gini_regional_percapita} are defined across all academic fields, they can also be defined for each depth of the field as well. Combined with the HHI measures per region, and GINI coefficients globally, the metrics give a sense of concentration and inequality of academic capital per geographical region, sliced by academic fields.

A healthy system would be indicated by different regions of the globe contributing specialized knowledge to the global pool while retaining sufficient cross-pollination between the academic fields. 
An unhealthy system would be indicated by a concentration of academic capital in certain geographies across multiple academic fields, implying no specialization, while certain other regions do not contribute to the knowledge pool at all. 
Also, greater collaboration between specialized geographical locations would lead to greater contributions in academic capital made by that region, to domains out of its specialization.

\subsection{Volatility} 

A lower volatility in FMP indicates a greater certainty in estimating the cost of quality control, whether through peer reviews, replications, or otherwise. A healthy system is expected to develop, eventually, better methods that help in estimating these costs, and hence a trend towards lower volatility is expected. 

Consider $\Bar{s}_p(t)$ be the global peer review fair market price. Then, volatility in peer reviewer marketplace over a time period $\{t-(n-1),\cdots,t-1,t\}$ can be computed as  
\begin{equation}
    \text{Volatility}_{p,n} = \sigma_{\Bar{s}_{p,n}} = \sqrt{\mathbb{E}[(\Bar{s}_{p}(t) - \mathbb{E}[\Bar{s}_{p}(t)])^2] \cdot n}
    \label{eq:global_volatility_fmp_reviews}
\end{equation}
Similarly, the volatilty for the replicators marketplace can be computed as 
\begin{equation}
    \text{Volatility}_{r,n} = \sigma_{\Bar{s}_{r,n}} = \sqrt{\mathbb{E}[(\Bar{s}_{r}(t) - \mathbb{E}[\Bar{s}_{r}(t)])^2] \cdot n}
    \label{eq:global_volatility_fmp_replications}
\end{equation}
\subsection{Global research efficiency} 
This is the efficiency of research considering all the manuscripts in the world, i.e., replace $\Pi$ with $\Theta$ in \cref{eq:funding_efficiency}, \cref{eq:time_efficiency}, i.e., the amount of academic capital produced by the globe as a whole normalized by the total amount of research spending (or the total time of ). At the globale scale, the normalization on funding could also be done by other global economic measures, e.g., GDP, PPP, etc.
\todo{P4: Mention Inflation adjusted funding efficiency -  AS}
\begin{equation}
    \text{Funding Efficiency}_{\Theta} = \epsilon_{\$, \Theta} = \frac{\ACEQ_{\Theta}}{\$_{\Theta}}
\end{equation}
\begin{equation}
    \text{GDP Efficiency}_{\Theta} = \epsilon_{\text{GDP}, \Theta} = \frac{\ACEQ_{\Theta}}{\$_{\Theta}} \cdot \text{GDP}_{\Theta}
\end{equation}
\begin{equation}
    \text{PPP Efficiency}_{\Theta} = \epsilon_{\text{PPP}, \Theta} = \frac{\ACEQ_{\Theta}}{\$_{\Theta}} \cdot \text{PPP}_{\Theta}
\end{equation}
\begin{equation}
    \text{Time Efficiency}_{\Theta} = \epsilon_{t, \Theta} = \frac{\ACEQ_{\Theta}}{\Delta t_{\Theta}}
\end{equation}
This metric tries to explain whether the increase in global academic capital is  due to increase in research spending, or due to the methodology of research becoming efficient as well. 
The global research system is expected to streamline its processes, enabled by better quantification of capital output and improvised quality control mechanisms. Hence, the research efficiency of the globe is expected to increase overall.

All the global research efficiency metrics described above can also be computed for smaller geographic regions $\theta_x \in \Theta$. This allows to investigate the research efficiency in a given geographical area, over a given period of time, so that the influence of policy and governance on academic output could also be measured.

\subsection{Transaction volume}
This is a measure of the total number of quality control actions (peer review or replications) that happens in a unit time.
This gives an estimate of the amount of quality control happening in the system.
Consider $\Pi_{\Delta t, \Theta}$ be the portfolio of all manuscripts produced in $\Delta t$ time period over the globe. Then, the transaction volume would be computed as the total number of peer reviewer and replicator nodes on those manuscripts. This can be easily computed by fetching $\hat{M} \in \Pi$ subset of manuscripts from the mask tensor $\mathscr{M}$'s peer reviewer and replicator slices, and taking a sum over the values.

\subsection{Time efficiency of quality control}
The amount of time peer review and replication take will be different across different academic fields, but should not have variability based on geographical location. 
Let $t_{p,d}(m)$ denote the time taken for peer review of manuscript $m$ in some academic field $d$, and $t_{r,d}(m)$ denote the time taken for replication of the manuscript. These values represent the sum total of all peer reviews and replications involved, if there were multiple peer reviewers or replicators.
Then, the time efficiency for quality control in a field $d$ is computed by taking a mean of the times over the subset $\hat{M}_D$ of manuscripts in $D$,
\begin{equation}
    \epsilon_{p,D} = \frac{1}{|\hat{M}_D|}\sum_{\hat{M}_D} t_{p,d}(m)
\end{equation}
\begin{equation}
    \epsilon_{r,D} = \frac{1}{|\hat{M}_D|}\sum_{\hat{M}_D} t_{r,d}(m)
\end{equation}

In a healthy system, we anticipate that the time efficiency of quality control will gradually increase, as better quality control is proportionately rewarded.

\subsection{Collection Subscription Ratio}
One way of measuring whether the academic impact of papers is only within it's own field, or outside of the academic field, is to look at the ratio of the number of collection subscribers to the number of authors in that collection. Recall from \cref{ssec:collections_and_journals} that a collection is defined by a set of tags $\hat{D}$ that could be from the 4D academic field tags or from filters like peer reviewed, replicated, particular authors, institutions, geographic regions, or time periods. Manuscripts with tags $D_m$ such that $\hat{D} \subseteq D_m$ automatically appear in that collection. The CSR for that collection is defined as: 
\begin{equation}
    \text{CSR}(\hat{D}) = \frac{\text{subscribers to } \hat{D}}{\text{authors in }\hat{D}}
\end{equation}
For an academic field, the higher this value, the more impactful the work in this academic field is said to be outside of that academic field. For a journal, the higher this value, the more influential that journal's work is said to be beyond the journal's own contributor community. 

\newpage
\section{Exploits \& Modifications}
\label{sec:exploits_&_modifications}
Any choice of how to count academic capital, there may be exploits or inadequacies with the $\ACEQ\space$ metric. Research impactfulness is a multidimensional quantity with different components and weightings for different people, much like how value generated in an economy also has many definitions beyond the raw currency value. Likewise with econometrics, overly focusing on any single one allows policy making to become biased in its optimization goals, potentially to detrimental effects. In practical circumstances, a suite of econometrics are used to portray a more complete picture of reality, and the same practice is encouraged for Liberata despite having a default, flagship way of calculating $\ACEQ$.
\todo{P3: Refresh what weighted citations are -  AS}
Liberata's academic capital calculations are designed to be modular, with information about how $\ACEQ$ is counted stored in the references graph (\cref{sec:references_graph}) and information about how $\ACEQ$ is to be split and attributed amongst contributors stored in the shares graph (\cref{sec:shares_graph}). This allows the swapping of either graph with newly defined graphs without affecting other graphs and metrics. To construct a suite of $\ACEQ$ variants to paint a more complete picture of academic impact, several proposed variants of the references graph $\mathbf{G_w}$ are given below, each designed to cover potential weaknesses or oversight regions of the others. 
\todo{DONE: P2: Make sure that these all modifications are well distinguished from what has already been done in literature -  AS}
\subsection{Time Modulated Weighted Citations}
\label{ssec:time_modulated_weighted_citations}
The first candidate modification replaces the weighted citations references graph matrix $\mathbf{G}_W$ with a time modulated weighted citations (TMWC) references graph matrix $\mathbf{G}_{WT}$ according to the operation below. This results in a version of capital assignment that allows larger duration (longer project time) works to give out more proportionately more capital than shorter duration works. 
This is an important distinction than other prior works \cite{Walker2007, Wang2013, Parolo2015} that take temporal component into consideration for citation weighting, as the proposed modification is more adaptive to different publication cycles of different disciplines and allows arbitrary time-dependent transformations.
\begin{equation}
    \mathbf{G}_{WT} = \mathbf{G}_W \circ 
    \left[
    \begin{array}{c}
        \mathbf{t_{3,1}} \\
        \hline
        \mathbf{t_{3,2}} \\
        \hline
        \vdots \\
        \hline
        \mathbf{t_{3,|M|}}
    \end{array}
    \right]
\end{equation}

Here, the average time to publication $t_3$ at the discipline tag $d_3$ level is averaged from prior literature of the same $d_3$ primary tag, and each row (corresponding to a manuscript) of $\mathbf{G}_W$ is scaled by multiplying the original value by the row manuscript's corresponding $t_{3}$ to yield $\mathbf{G}_{WT}$. While slightly harder to interpret from a given number, this version of capital assignment is a reasonable alternative to the simple weighted citations capital assignment method. For $\ACEQ$ to be counted in this way, a counter-incentive is needed to guard against artificial inflation of project time for more academic credit, and we propose the time efficiency in \cref{ssec:research_efficiency} metric for this. In addition, this modification, along with others, are not meant to be used in isolation and are instead meant to be used in unison to depict a more accurate picture of academic impact.

\subsection{Impact Modulated Weighted Citations}
\label{ssec:impact_modulated_weighted_citations}
A second candidate modification, called impact modulated weighted citations (IMWC) scales up the citation power of each article according to their impact, measured using the default definition of $\ACEQ$ in \cref{ACDefinition}. To ensure asymptotic convergence and to prevent distortions from long self citation chains, rows of the weighted citations references graph $\mathbf{G}_W$ are scaled by an impact modifier $\iota$ with a logarithmically decaying construction. This construction is related to recursive prestige-based citation weighting methods such as PageRank and Pinski–Narin influence models \cite{Page1999,Pinski1976,Chen2007}, but differs in explicitly modulating the citation adjacency matrix using a nonlinear, log-damped impact transformation, and in employing a truncated iterative scheme.
\begin{equation}
    \iota(m \in M) = log_2(\ACEQ_m+1)
\end{equation}
\begin{equation}
    \mathbf{G}_{W\iota} = \mathbf{G}_W \circ 
    \left[
    \begin{array}{c}
        \boldsymbol{\iota_{1}} \\
        \hline
        \boldsymbol{\iota_{2}} \\
        \hline
        \vdots \\
        \hline
        \boldsymbol{\iota_{|M|}}
    \end{array}
    \right]
\end{equation}
This alternative to assigning academic capital scales up the citation power of works logarithmically according to their own impactfulness, rewarding works that lead to impactful works more. Importantly, the causal chain is kept short by the logarithmic nature of this impact factor definition, preventing large distortions to second degree ancestral works and beyond. This definition also implies an iterative solver, but the iterations can be truncated after four steps due to the fourth logarithm being so insignificantly small that we practically reach the asymptotic value after three or four steps. 

\subsection{Author Cosine Similarity Discounting}
A third candidate modification is to discount $\ACEQ$ going from authors to themselves, also known as self citation. In this method, we construct a matrix of author cosine similarity modifiers (ACSM):

\begin{equation}
    \Phi =
    \begin{bmatrix}
    0         & \phi_{1,2} & \cdots      & \phi_{1,n-1} & \phi_{1,n} \\
    0         & 0          & \cdots      & \phi_{2,n-1} & \phi_{2,n} \\
    \vdots    & \vdots     & \ddots      & \vdots       & \vdots     \\
    0         & 0          & \cdots      & 0            & \phi_{n-1,n} \\
    0         & 0          & \cdots      & 0            & 0
    \end{bmatrix}
\end{equation}
with each element defined as:
\begin{equation}
    \phi_{k,l}({\bar{s}_k',\bar{s}'_l}) =
    \begin{cases}
    \bar{s}_k' \cdot \bar{s}_l' & \text{if } m_l \text{ cites } m_k \\
    0  & \text{otherwise}
    \end{cases}
\end{equation}
Here, $k$ and $l$ are two indices corresponding to two manuscripts $m_k$ and $m_l$, and $\bar{s}_k'$ and $\bar{s}_l'$ are the extended shares split vectors for $m_k$ and $m_l$ respectively, constructed via the process below. 
\begin{enumerate}
    \item Take the union of authors in $m_k$ and $m_l$, $A_{k,l} = A_k \bigcup A_l$
    \item Assign each $a \in A_{k,l}$ a unique index $n$.
    \item Set the length of $\bar{s}_k'$ and $\bar{s}_l'$ to be of size $n$.
    \item Assign index $n$ of $\bar{s}_k'$ and $\bar{s}_l'$ to be the shares that author $a_n$ holds in $m_k$ and $m_l$ respectively.
\end{enumerate}
The weighted citations references graph $\mathbf{G_{W}}$ is then multiplied element wise by $\mathbb{1}-\Phi$ to yield the ACSM references graph, where $\mathbb{1}$ is the ones matrix:
\begin{equation}
    \mathbf{G}_{W\Phi}= \mathbf{G}_W \circ (\mathbb{1}-\Phi)
\end{equation}
This modification attempts to discourage self citing by discounting citations coming from works by the similarity of the author list share distributions. This modification generalizes the idea of fractional self-citation proposed by \cite{Schubert2006} by using author-share vectors to compute a continuous edge-level discount, so that citations are penalized in proportion to the similarity of contribution distributions across the citing and cited manuscripts. This modification, like the others, is not meant to be used in isolation, but instead is meant to give independent information in conjunction with other modifications.

\subsection{Alternative Modifications}
The modular nature of Liberata's references graph also allows for the linear modifications above, which are commutative and associative, to be chained together in any order to form compound modifications.

The modifications above are expected to be a small subset of possible and sensible modifications to how $\ACEQ$ is counted. As other modifications are developed, they too can be swapped in or combined with the ones defined above. Such modifications might try to take into account information about intellectual property generated rom the research, or attempt to quantify downstream economic impacts, or even propitiatory custom constructs similar to Altmetrics. \cite{Priem2012} It is hoped that a suite of future variants created by the academic community will illuminate the landscape of academic contributions if/when the Liberata system sees increasing adoption.

\newpage
\section{Conclusion}

This paper is intended as a reference catalog and an argument for continuous valued market based metrics. Its central claim is that replacing discrete authorship with continuous contribution shares, and unweighted citations with normalized and corrected weighted ones, unlocks more accurate and insightful scientometrics, while a coupled academic marketplace for trading credit for services unlocks more reliable and robust academic quality control.

Central to the construction of Liberata's scientometrics are the shares graph $G_S$ and the references graph $G_W$, which together compose the capital graph $G_\ACEQ$ and support a broad and deep set of capabilities. Graph-theoretic properties of the shares network, including its spectral structure, spanning tree counts, and two-step compositions, characterize collaboration topology, global connectivity, and potential collusion between authors and their quality controllers. 

Portfolio metrics, well defined over any collection of manuscripts and contributors, provide measures of impact,
concentration, collaboration structure, and research efficiency that extend naturally from individuals, to institutions, fields, and geographic regions resolving attribution ambiguities that render noisy comparisons in the current system.

Market metrics derived from share transactions yield field specific value of quality control services, risk premiums of entities, and risk-adjusted performance measures that have no analogy in conventional metrics. Modular correction factors to the references graph, linearly composable, provide a flexible system for comparing and counting how impact propagates through the citation network under different impact definitions.

Liberata's publishing system offers a way to correct incentive alignment problems between authors, quality controllers, and the broader academic community. Liberata's scientometrics offers a way to detect each type of conventionally recognized and exercised exploitative strategy, enabling such a platform and many other rich scientometric applications to be developed in future work. 

\newpage
\section{Notation}
\subsection{Node Classes}
\begin{description}[leftmargin=4cm, style=nextline]
    \item[$m \in \hat{M} \subset M$] Manuscripts
    \item[$a \in \hat{A} \subset A$] Authors
    \item[$p \in \hat{P} \subset P$] Peer reviewers
    \item[$r \in \hat{R} \subset R$] Replicators
    \item[$c \in \hat{C} \subset C$] Contributors
    \item[$i \in \hat{I} \subset I$] Institutions
    \item[$t \in \hat{T} \subset T$] Timestamps
    \item[$\theta \in \hat{\Theta} \subset \Theta$] Geographic regions
\end{description}

\subsection{Intermediates}
\begin{description}[leftmargin=4cm, style=nextline]
    \item[$x, y, z$] Various indexing
    \item[$e$] Euler's constant
    \item[$\lambda, \Lambda$] Eigenvalue, Diagonal Matrix of Eigenvalues
    \item[$k \in K, l \in L,  n \in N$] Counters
    \item[$\mathbb{R}$] Real numbers
    \item[$\mathbb{N}$] Natural numbers
    \item[$\mathbb{Z}$] Integers
    \item[$\mathbb{E}$] Expected value
    \item[$b$] Arbitrary parameter
    \item[$\bar{s}$] Bar indicates vector of the object type underneath the bar
\end{description}

\subsection{Graphs \& Matrices}
\begin{description}[leftmargin=4cm, style=nextline]
    \item[$u \in \hat{U} \subset U$] Unweighted citations
    \item[$w \in \hat{W} \subset W$] Weighted citations
    \item[$G$] Graph
    \item[$\mathbf{G}$] Adjacency matrix of graph
    \item[$\mathbf{B}$] Block of a matrix
    \item[$V$] Set of vertices
    \item[$E$] Set of edges
    \item[$\mathbf{S}$] Matrix/submatrix of shares
    \item[$\mathbf{v}$] Unit vectors
    \item[$\iota$] Impact modifiers
    \item[$\phi$] Cosine similarity modifier
\end{description}

\subsection{Capital \& Markets}
\begin{description}[leftmargin=4cm, style=nextline]
    \item[$s \in \hat{S} \subset S$] Shares
    \item[$\ACEQ$] Academic capital
    \item[$\Delta \ACEQ$] Return
    \item[$\mu$] Expected returns
    \item[$\sigma$] Volatility
    \item[$\gamma$] Risk asymmetry
    \item[$\epsilon$] Efficiency
    \item[$\Bar{s}_p, \Bar{s}_r \in \Bar{S}$] Fair market prices
    \item[$\psi$] Risk premium
    \item[$\nu$] Utility
    \item [$\alpha$] Excess returns (outperformance)
    \item [$\beta$] Market sensitivity
    \item[$d_{1,2,3,4} \in D_{1,2,3,4}$] Academic domain, department, discipline, or direction
    \item[$\text{ARC}$] Academic returns to capital ratio
    \item[$\text{HHI}$] Herfindahl–Hirschman Index
    \item[$\text{Gini}$] Gini Index
    \item[$\pi \in \Pi$] Portfolio 
\end{description}

\newpage
\bibliographystyle{unsrtnat}
\bibliography{references}  

\end{document}